\documentclass[lettersize,journal]{IEEEtran}
\usepackage{amsmath,amsfonts}
\usepackage{array}
\usepackage{graphics}    % figure setting
\usepackage{subfigure}     % caption setting
\usepackage{textcomp}
\usepackage{stfloats}
\usepackage{url}
\usepackage{verbatim}
\usepackage{graphicx}
\usepackage{cite}
\usepackage{multirow}
\usepackage{array}
\usepackage{multicol}
\usepackage{caption}
\usepackage{subcaption}
\usepackage[ruled,linesnumbered]{algorithm2e}
\usepackage[noend]{algpseudocode}
\usepackage{tabularx}
\usepackage{multirow}
\usepackage{bbding}
\makeatletter
\let\NAT@parse\undefined
\makeatother
\usepackage{hyperref}  %hyperref still needs to be put at the end!
\usepackage{threeparttable}    %% table note
\usepackage{makecell}
\usepackage{float}
\begin{document}

\title{SISSA: Real-time Monitoring of Hardware Functional Safety and Cybersecurity with In-vehicle SOME/IP Ethernet Traffic}

\author{ Qi Liu, Xingyu Li, Ke Sun, Yufeng Li*, Yanchen Liu*
\thanks{
* Corresponding author.

Qi Liu, Xingyu Li, Ke Sun, Yufeng Li, Yanchen Liu
 are with the School of Computer Engineering and Science, Shanghai University, Shanghai 200444, China

 Y. Li is also with the Purple Mountain Laboratories, Nanjing 211111, China (e-mail: liyufeng\_shu@shu.edu.cn)
}}
        % <-this % stops a space

%\markboth{Journal of \LaTeX\ Class Files,~*.~*, No.~*, *~*}%
%\markboth{}%
%{Shell \MakeLowercase{\textit{et al.}}: A Sample Article %Using IEEEtran.cls for IEEE Journals}

%\thanks{This paper was produced by the IEEE Publication Technology Group. They are in Piscataway, NJ.

%Manuscript received April 19, 2021; revised August 16, 2021.}

%\thanks{This paper was produced by the IEEE Publication Technology Group. They are in Piscataway, NJ.
%Manuscript received April 19, 2021; revised August 16, 2021.}



\maketitle

\begin{abstract}
%Identifying malfunctions and cyberattacks in Electronic Control Units (ECUs) is vital for the development of connected automated vehicles (CAVs).
 Scalable service-Oriented Middleware over IP (SOME/IP) is an Ethernet communication standard protocol in the Automotive Open System Architecture (AUTOSAR), promoting ECU-to-ECU communication over the IP stack. However, SOME/IP lacks a robust security architecture, making it susceptible to potential attacks. Besides, random hardware failure of ECU will disrupt SOME/IP communication. In this paper, we propose SISSA, a SOME/IP communication traffic-based approach for modeling and analyzing in-vehicle functional safety and cyber security. Specifically, 
SISSA models hardware failures with the Weibull distribution and addresses five potential attacks on SOME/IP communication, including Distributed Denial-of-Services, Man-in-the-Middle, and abnormal communication processes, assuming a malicious user accesses the in-vehicle network. Subsequently, 
SISSA designs a series of deep learning models with various backbones to extract features from SOME/IP sessions among ECUs. We adopt residual self-attention to accelerate the model's convergence and enhance detection accuracy, determining whether an ECU is under attack, facing functional failure, or operating normally. Additionally, we have created and annotated a dataset encompassing various classes, including indicators of attack, functionality, and normalcy. This contribution is noteworthy due to the scarcity of publicly accessible datasets with such characteristics.
Extensive experimental results show the effectiveness and efficiency of SISSA.

\end{abstract}


\section{Introduction}
\IEEEPARstart{R}{cent} years have witnessed rapid advancements in the connected and automated vehicles (CAVs), whose key features include the increasing number of Electronic Control Units (ECUs). ECUs play a pivotal role in managing various vehicle functions, such as the throttle, brake, and steering. To fulfill these functions, there is a growing demand for high and flexible communication bandwidths, often realized through Automotive Ethernet (e.g., BroadR-Reach) and IP-based communication \cite{iorio2020protecting}. SOME/IP has been specifically designed to meet automotive requirements, including compatible with AUTOSAR and ensuring fast response times \cite{autosar_someip}. It supports devices of various sizes and operating systems, effectively replacing traditional vehicle network technologies such as Control Area Network (CAN) bus, Media Oriented Systems Transport (MOST), or FlexRay \cite{zelle2021analyzing}.  

Despite these advantages, ECUs relying on SOME/IP communication still encounter challenges during operation, including random functional failures (safety) and malicious cyberattack (security). ECUs typically constitute entities wherein hardware and software collaborate closely. Inevitably, hardware components encounter random hardware failures during their lifecycle, thereby impacting the normal functioning of the ECU. Furthermore, SOME/IP specifications are devoid of security measures directly integrated into the protocol \cite{iorio2020securing}. As a result, security vulnerabilities in SOME/IP deployments introduce novel attack vectors that can be exploited to target victim vehicle. Research has demonstrated that numerous commercial vehicles display vulnerabilities in network stack implementations, particularly in Bluetooth, Wi-Fi, and 4G. Consequently, this exposes the potential for malicious entities to remotely access the internal systems of cars \cite{miller2015remote}.

To analyze the random functional failures and malicious cyberattacks of automated vehicle, existing countermeasures can be divided into two categories: risk assessment, and feature analysis. 
Hazard Analysis and Risk Assessment (HARA) and Threat Analysis and Risk Assessment (TARA) are fundamental elements of risk assessment in the context of CAVs. These processes play pivotal roles as essential stages within the ISO 26262 \cite{ISO-26262} and ISO/SAE 21434 standard \cite{international2021iso}, respectively. Typically, HARA employs failure-based methods such as Fault Tree \cite{ruijters2015fault} and Failure Modes and Effects Analysis \cite{asq_fmea}, focusing on potential faults and their impact on system safety. Additionally, system-based approaches like HAZOP \cite{signoret2021hazard} and System Theoretic Process Analysis \cite{leveson2016engineering} comprehensively assess the system's structure, operation, and environmental factors to evaluate overall safety. TARA, including methods like FMVEA \cite{schmittner2014fmvea}, STPA-sec \cite{young2017system}, and VeRA \cite{cui2020vera}, involves the analysis of potential threats and the formulation of appropriate mitigation strategies.

Automotive employs a variety of features, such as the entropy of In-vehicle Network \cite{muter2011entropy}, message intervals \cite{miller2013adventures}, message correlation/consistency \cite{muter2010structured}, ECU fingerprints \cite{cho2016fingerprinting}, vehicle voltage \cite{xun2021vehicleeids}, communication characteristics \cite{choi2018voltageids}  \cite{alkhatib2021some} \cite{luo2023multi}\cite{alkhatib2023said}, etc., serving as crucial indicators in identifying deviations from normal behavior within the vehicle network. Statistical methods ( such as frequency statistics) and machine learning techniques (Convolutional Neural Network, Long Short-Term Memory) leverage these features to effectively detect and predict abnormal patterns. 

However, above functional failures and malicious cyberattacks analyzing methods have the following limitations in the context of SOME/IP communication. First, risk assessment methods, such as HARA/TARA, is explored in concept phase but focuses on changes at the component and system levels, does not take into account variations in in-vehicle communication protocol data packets. Second, feature learning encounters practical challenges, suffering from time-consuming parameter adjustments, and demanding considerable computation power. Third, SOME/IP's network traffic is considerably more dynamic and unpredictable than classic automotive protocols like CAN. Establishing a baseline for normal operation and detecting network flow anomalies becomes challenging. 

To overcome the above limitations, we propose a real-time in-vehicle safety and security monitoring approach using SOME/IP traffics (SISSA). Firstly, We model random hardware failures and classic attacks on ECUs while considering SOME/IP communication traffic. We employ Weibull distribution to model random hardware failures of ECUs. Given unauthorized access to the in-vehicle network, classic attacks on SOME/IP communication, including Distributed Denial-of-Services, Man-in-the-Middle, and abnormal communication processes, are identified. 
Then, we propose three deep learning-based models to extract features from SOME/IP packets. The adoption of residual self-attention accelerates model convergence and improves detection accuracy. To establish a baseline for normal communication evaluation, we create and annotate a dateset comprising diverse classes, encompassing indicators of attack, functionality, and normalcy. This contribution stands out because there are very few datasets available to the public that possess these particular characteristics. 

The main contributions of this paper can be summarized as follow:

\begin{itemize}
  \item We propose a novel approach named SISSA to monitor in-vehicle safety and security using SOME/IP Ethernet traffics. Our contribution involves the modeling of functional failure and cyberattack scenarios in the communication of ECUs through SOME/IP traffic. Additionally, we propose three deep learning-based models integrated with residual self-attention to determine the state of ECUs. 
  \item We propose a  SOME/IP data generation method for establishing a baseline for normal communication evaluation. It encompasses the classes of attack, functional failure and normal. Both code and datasets can be accessed at https://github.com/jamesnulliu/SISSA.  
  \item We conducted extensive experimental assessments of SISSA, yielding an average F1-score of 99.8\% for malfunction identification and a flawless 100\% average F1-score for cyberattack detection. The detection speed and model overhead satisfy the communication requirements of automotive networks and  the computational resource demands of ECUs.
\end{itemize}

The rest of the paper is organized as follows. The SOME/IP architecture and communication model are introduced in Section \ref{preliminary}. The state-of-the-art safety and security monitoring methods of CAVs are reviewed in Section \ref{Related works}. Section \ref{Methodology} presents the specific design of our safety and security monitoring model, while the performance evaluation is presented in Section \ref{Evaluation}. Finally, Section \ref{Conclusion} concludes the paper. 

\section{Preliminary}
\label{preliminary}
\subsection{SOME/IP Protocol}
\label{some/ip protocol}
The SOME/IP protocol serves as a scalable and service-oriented middleware seamlessly integrated within the IP architecture. It stands out as a distinctive member within the TCP/IP protocol suite. Initially proposed by BMW in 2011, the protocol was subsequently integrated into the AUTOSAR standard in 2014 \cite{gehrmann2020intrusion} \cite{autosar_protocol}. It distinguishes itself from other in-vehicle communication protocols, such as CAN and LIN, by offering support for remote procedure calls (RPC), event notifications, and the underlying serialization/wire format.

RPC facilitates the retrieval of information from an ECU server by an ECU client. Notifications describe a general Publish/Subscribe mechanism, which effectively manages the publish and subscribe relationship of a service \cite{autosar_someip}. Serialization describes the representation of data in the form of Protocol Data Units (PDUs) as payloads of UDP or TCP messages for transmission over IP-based vehicular networks. Furthermore, one of the most noteworthy features provided by SOME/IP is service discovery (SD), dynamically advertising the availability of various services and implementing Publish/Subscribe management \cite{autosar_somediscovery}.

\begin{figure}[!t]
	\centering
	\includegraphics[width=3.5in]{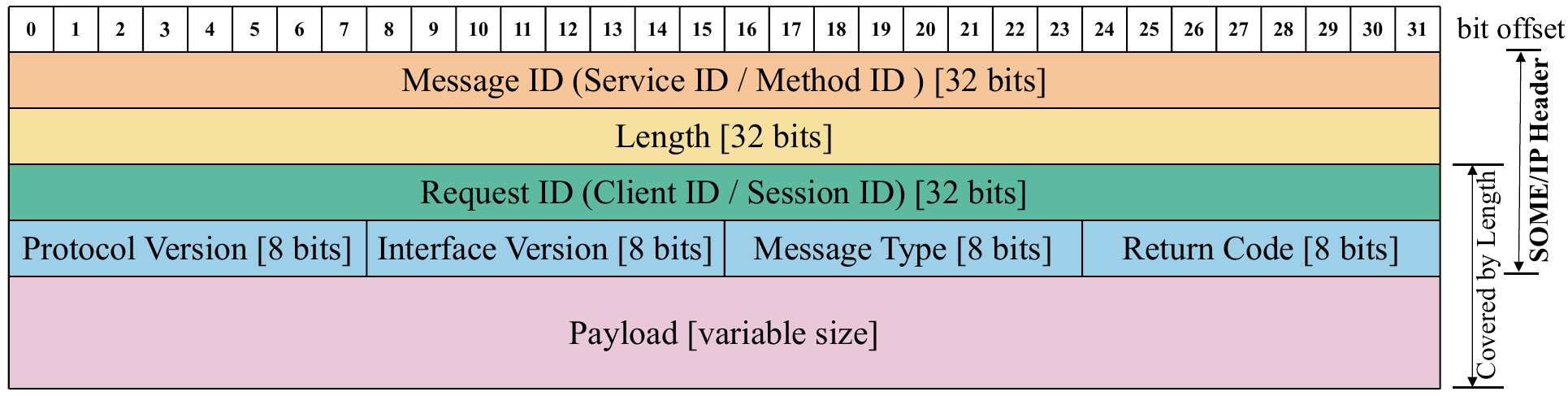}
	\caption{SOME/IP packet}
	\label{fig: packet}
\end{figure}

Figure \ref{fig: packet} shows the header frame format of SOME/IP protocol. According to AUTOSAR standard,  SOME/IP header consists of the following fields:

\begin{itemize}
	\item Message ID (Service ID / Method ID)[32 bits]: A message ID is a 32-bit identifier composed of a 16-bit Service ID and a 16-bit Method ID. It is used to identify an RPC call to an application or to identify an event.
	\item Length[32 bits]: The length field contains the length in bytes from the Request / Client ID to the end of the SOME/IP message. 
	\item Request ID (Client ID / Session ID) [32 bits]: Request ID is 32-bit and consists of a 16-bit Client ID and a 16-bit Session ID.
	\item Protocol Version [8 bits]: Protocol Version identifies the used SOME/IP header format (excluding the payload format) and is 8 bytes in length.
	\item Interface Version [8 bits]: Interface Version contains the main field interface version, which consists of 8 bytes.
	\item Message Type [8 bits]: Message Type is used to distinguish between different types of packets. Normally, there are ten different message types for SOME/IP protocols.
	\item Return Code [8 bits]: Return Code is used to indicate whether it is a request or not Processing was successful.
	\item Payload[variable size]: Payload represents the information content to be delivered.
\end{itemize}

\subsection{Communication model}
\begin{figure}[!t]
	\centering
	\includegraphics[width=3.5in]{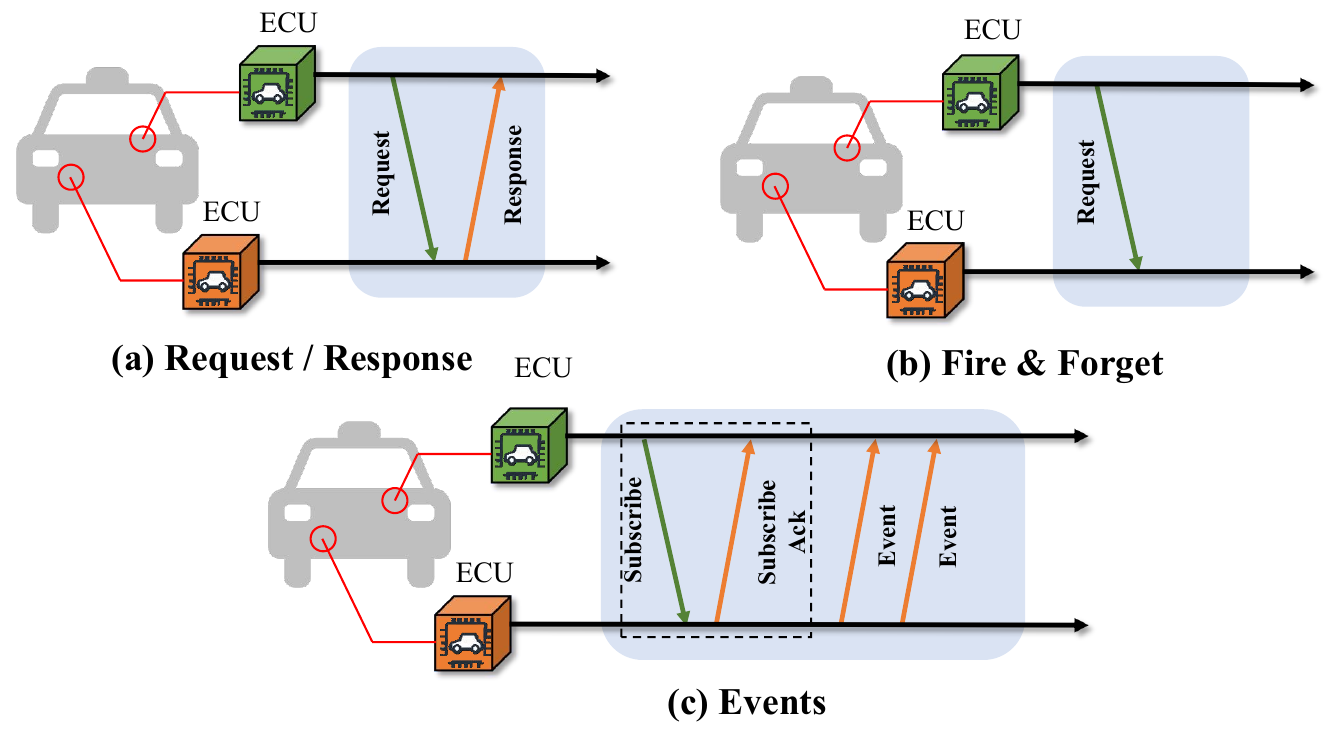}
	\caption{Three communication models in SOME/IP}
	\label{fig: communication mode}
\end{figure}

In this section, we consider the communication model of SOME/IP. The SOME/IP protocol defines three main communication models as shown in Figure \ref{fig: communication mode}.

\textbf{Request/Response}: The Request/Response pattern is one of the most common communication patterns, where a request is initiated by one communication partner (client) and responded by another communication partner (server). Request/Response communication is essentially a RPC that consists of a request and a response \citenum{autosar_somediscovery}. Specifically, a typical request bearing the message type 0x00 anticipates a singular response denoted by the message type 0x80. 

\textbf{Fire \& Forget}: This communication is a RPC that consists only of a request message. In the Fire \& Forget communication pattern, the request initiated by the message sender does not necessitate a response message. The request message type associated with this pattern is REQUEST\_NO\_RETURN.

\textbf{Events}: This communication pattern entails a client sending a subscription request to access a service. In certain scenarios, the server may dispatch an event to the client, which means the client that sent the subscription request will receive the service, and the Publish/Subscribe is determined by SOME/IP-SD. Various sending strategies are employed for different events, including periodic updates from the server to the client or immediate notifications upon any change in a value. It is noteworthy that, for both subscription and event messages, the designated message type is NOTIFICATION. Importantly, when the server transmits an event to the client, the client is not required to provide a response.

\section{Related works}
\label{Related works}
In this Section, we classify the state-of-the-art safety and security analyzing methods of CAVs from three perspectives: component-based, system-based, and communication packet-based. This classification ensures a comprehensive approach, covering individual elements, their integrated functionality, and the safe/secure data exchange. It is noted that HARA/TARA is explored to analyze the safety and security of components and systems, respectively. Conventional features within the in-vehicle communication network are designed to identify abnormal behavior. Subsequently, we delve into a detailed review of risk assessment-based and feature learning-based analysis methods as follows.  

\subsection{Risk assessment based}
HARA is a pivotal part developed within functional safety standard ISO 26262. It systematically identifies and assesses potential hazards and risks within Electrical/Electronic systems, offering guidance during the design and development phases to ensure that vehicles comply with rigorous functional safety requirements throughout their entire lifecycle. HARA employs failure-based and system-based methods.  For instance, Fault Tree \cite{ruijters2015fault} analyzes and represents the logical relationships among various potential faults and their contributing events within a system. Failure Modes and Effects Analysis (FMEA) \cite{asq_fmea} evaluate sand prioritizes potential failure modes in a system, identifying their causes and effects, and implementing preventive measures to enhance safety. Besides, HAZOP (Hazard and Operability Study) \cite{signoret2021hazard} systematically identify and assess potential hazards and operability issues by examining deviations from the intended design conditions. STPA \cite{leveson2016engineering} mitigates hazardous states and their corresponding causal factors
within complex systems by employing system theory principles.

Aligned with vehicle cybersecurity standard ISO/SAE 21434, TARA ensures effective defense mechanisms and significant cost savings in the early stages of vehicle development. It systematically analyzes potential threats to vehicle systems, determining hierarchical defense strategies and implementing corresponding mitigations based on identified threats. TARA employs formula-based methods and model-based methods \cite{luo2021threat}. Formula-based methods include EVITA \cite{henniger2009securing}, HEAVENS \cite{sae_j3061}, and SARA \cite{monteuuis2018sara}, etc., mainly through tables, texts, or formulas. Model-based methods like PASTA \cite{kim2020internet}, STPA-sec \cite{young2017system}, and Attack Tree \cite{ren2011novel}, etc., modeling and analyzing the threats and risks of the system through data flow diagrams, graphs, and tree models. Formula-based TARA methods, due to their maturity and user-friendly nature, enjoy broader adoption and usage, especially among users with limited security experience. Formula-based TARA methods are more mature and more convenient for users without too much security experience. As a result, they are more widely spread and used. While model-based methods are more complex and therefore more difficult to understand and use. 

\begin{table*}
	\centering
	\caption{Comparison with existing vehicular SOME/IP -based safety and security analysis methods. The criteria include: the exploited communication model, the type of detection algorithms, the analysis target, the open-source data and code, and the real-time capacity. RB and ML indicate the rule-based, machine-learning based.}
	\label{table:Comprison with existing}
	\begin{tabular}{c|ccc|c|cc|c|c}
		\hline
		\multirow{2}{*}{\textbf{Literature}}                & \multicolumn{3}{c|}{\textbf{Communication models}}                    & \multirow{2}{*}{\textbf{Type}}                    & \multicolumn{2}{c|}{\textbf{Targets}}                                                                & \multirow{2}{*}{\textbf{Oa}} & \multirow{2}{*}{\textbf{Re}} \\ \cline{2-4} \cline{6-7}
		& \textbf{Req/Res}          & \textbf{F\&F}              & \textbf{E}                  &                                          & \textbf{Sa}                 & \textbf{Se}                                           &                     &                     \\ \hline
		\multirow{2}{*}{\cite{iorio2020securing}}       & \multirow{2}{*}{\checkmark} & \multirow{2}{*}{$\times$} & \multirow{2}{*}{$\times$} & \multirow{2}{*}{Cryptography}            & \multirow{2}{*}{$\times$} & Replay, eavesdrop, modify and                & \multirow{2}{*}{\checkmark}  & \multirow{2}{*}{$\times$}  \\
		&                    &                    &                    &                                          &                    & drop arbitrary messages                      &                     &                     \\
		\multirow{2}{*}{\cite{alkhatib2021some}}    & \multirow{2}{*}{\checkmark} & \multirow{2}{*}{\checkmark} & \multirow{2}{*}{\checkmark} & \multirow{2}{*}{ML (RNN)}                & \multirow{2}{*}{$\times$} & Request without Response, Error on Error,    & \multirow{2}{*}{\checkmark}  & \multirow{2}{*}{$\times$}  \\
		&                    &                    &                    &                                          &                    & Response without Request, Error on Event    &                     &                     \\
		\cite{ma2022authentication}                          & \multirow{1}{*}{\checkmark}                  & \multirow{1}{*}{\checkmark}                 & \multirow{1}{*}{\checkmark}                 & Authentication                           & ×                  & Replay, eavesdrop, MITM, Masquerading attack & \multirow{1}{*}{\checkmark}                 & ×                   \\
		\multirow{2}{*}{\cite{casparsen2022closing}} & \multirow{2}{*}{\checkmark} & \multirow{2}{*}{\checkmark} & \multirow{2}{*}{\checkmark} & \multirow{2}{*}{RB}                      & \multirow{2}{*}{$\times$} & RPC Flooding, False Notify,                  & \multirow{2}{*}{$\times$}  & \multirow{2}{*}{$\times$}  \\
		&                    &                    &                    &                                          &                    & Session Stealing, Stop Offer Attack          &                     &                     \\
		\multirow{2}{*}{\cite{luo2023multi}}        & \multirow{2}{*}{\checkmark} & \multirow{2}{*}{\checkmark} & \multirow{2}{*}{\checkmark} & \multirow{2}{*}{RB and ML (GRU) }        & \multirow{2}{*}{$\times$} & fuzzy, Abnormal Communication  Process,      & \multirow{2}{*}{$\times$}  & \multirow{2}{*}{\checkmark}  \\
		&                    &                    &                    &                                          &                    &  spoof, DDoS,  Unauthorized Operation         &                     &                     \\
		\multirow{2}{*}{\cite{gehrmann2020intrusion}}  & \multirow{2}{*}{\checkmark} & \multirow{2}{*}{$\times$} & \multirow{2}{*}{\checkmark} & \multirow{2}{*}{Rule and Signature}      & \multirow{2}{*}{$\times$} & Detecting function invocation,                & \multirow{2}{*}{$\times$}  & \multirow{2}{*}{$\times$}  \\
		&                    &                    &                    &                                          &                    & no specific attacks                          &                     &                     \\
		\cite{koyama2022some}                       & \multirow{1}{*}{\checkmark}                 & \multirow{1}{*}{$\times$}                 & \multirow{1}{*}{\checkmark}                  & whitelist-based                          & \multirow{1}{*}{$\times$}                  & Replay, MITM                                & \multirow{1}{*}{$\times$}                   & \multirow{1}{*}{$\times$}                   \\
		\cite{zelle2021analyzing}                        & \multirow{1}{*}{\checkmark}                  & \multirow{1}{*}{$\times$}                  & \multirow{1}{*}{$\times$}                  & Cryptography                             & \multirow{1}{*}{$\times$}                 & MITM                                         & \multirow{1}{*}{$\times$}                   & \multirow{1}{*}{$\times$}                   \\
		\multirow{2}{*}{\cite{alkhatib2023said}}   & \multirow{2}{*}{\checkmark} & \multirow{2}{*}{\checkmark} & \multirow{2}{*}{\checkmark} & \multirow{2}{*}{ML (Transformer)}                      & \multirow{2}{*}{$\times$} & Request without Response, Error on Error,    & \multirow{2}{*}{\checkmark}  & \multirow{2}{*}{\checkmark}  \\
		&                    &                    &                    &                                          &                    & Response without Request, Error on Event    &                     &                     \\
		\cite{lee2023protecting}                         & \multirow{1}{*}{\checkmark}                  & \multirow{1}{*}{$\times$}                    & \multirow{1}{*}{$\times$}                & Authentication                           & \multirow{1}{*}{$\times$}                  & MITM, replay,  DDos, spoof                   & \multirow{1}{*}{\checkmark}                   & \multirow{1}{*}{$\times$}                   \\ \hline
		\multirow{2}{*}{\textbf{SISSA}}                  & \multirow{2}{*}{\textbf{\checkmark}} & \multirow{2}{*}{\textbf{\checkmark}} & \multirow{2}{*}{\textbf{\checkmark}} & \multicolumn{1}{c|}{\textbf{RB and ML }}          & \multirow{2}{*}{\textbf{\checkmark}} & \textbf{DDos, MITM}                                    & \multirow{2}{*}{\textbf{\checkmark}}  & \multirow{2}{*}{\textbf{\checkmark}}  \\
		&                    &                    &                    & \multicolumn{1}{l|}{\textbf{(CNN, RNN and LSTM)}} &                    & \textbf{abnormal communication}                       &                     &                     \\ \hline
	\end{tabular}

    \begin{tablenotes}
	\footnotesize
	%\item[*] this is the ....  %此处加入注释*信息
	%\item[**] my website is ... %此处加入注释**信息
	\item Req/Res: Request/Request; F\&F: Fire\&Forget; E: Event; Sa: Safety; Se: Security; Oa: Open access; Re: Real time; MITM: Man-in-the-Middle ; DDos: Distributed Denial-of-Services; 
    \end{tablenotes}

\end{table*}

\subsection{Feature learning based}
The vehicular abnormal detection aims to detect anomalies through exploiting in-vehicle features. It’s noteworthy that abnormal detection on the CAN bus has advanced significantly, supported by ample related research literature. Existing CAN bus abnormal detection can be divided into ID-based methods, payload-based methods, signal-based methods, and indirect features-based methods \cite{zhang2023bit}. ID-based methods monitor the entropy \cite{muter2011entropy}, frequency \cite{taylor2015frequency}, and time arrival interval \cite{song2016intrusion} of frame ID, which will trigger an alarm when they exceed the pre-designed threshold. Payload-based methods usually adopt machine learning model, such as deep neural networks \cite{javed2021canintelliids}, generative adversarial network \cite{seo2018gids} to learn the feature of frame payload, determining abnormal pattern. Signal-based methods analyze vehicle sensor data, such as speed, steering angle to detect anomalies \cite{he2020exploring}\cite{xue2022said}. Additionally, indirect features, such as ECU fingerprints \cite{cho2016fingerprinting}, are used to locate the compromised ECU. However, above methods are suitable for stable CAN message sequences, but not applicable to the dynamic and unpredictable nature of SOME/IP traffic.

While feature learning for SOME/IP is in its early stages, with limited research in this area. Based on the criteria: the exploited communication model, the type of analysis algorithms, the analysis target, the open-source data and code, and the real-time capacity, we review recent SOME/IP-based safety and security analysis methods in table \ref{table:Comprison with existing}. Cryptography and Authentication techniques are adopted to secure SOME/IP communication between ECUs \cite{zelle2021analyzing} \cite{iorio2020securing} \cite{ma2022authentication}\cite{lee2023protecting}. However, the application of encryption authentication methods tends to introduce increased communication complexity, elevated computational overhead, and additional latency. Koyama et.al \cite{koyama2022some} propose to real-timely detect anomalies through whitelist-based ways. Notably, their focus is primarily on mitigating replay attacks and man-in-the-middle attacks. To enhance detection accuracy and broaden the spectrum of detected attack types, machine learning models, such as RNN, GRU, and Vaswani’s transformer model, are employed to learn SOME/IP communication packet features and determine whether anomalies exist in the target ECU \cite{alkhatib2021some} \cite{luo2023multi}\cite{alkhatib2023said}. However, machine learning-based methods are specifically employed for detecting cyberattacks and are unable to identify malfunction within the ECU. 

Table \ref{table:Comprison with existing} summarizes the major differences between SISSA and the existing vehicular SOME/IP abnormal detection methods. The proposed SISSA can real-time monitor both functional failures and cyberattacks in the target ECU, effectively handling various communication models, including Request/Response, Fire and Forget, and Events. Additionally, our data and code are open-source, facilitating researchers in pursuing their studies.

\section{The proposed safety and security monitoring model}
\label{Methodology}

\begin{figure*}[!t]
\centering
\includegraphics[width=7.1in]{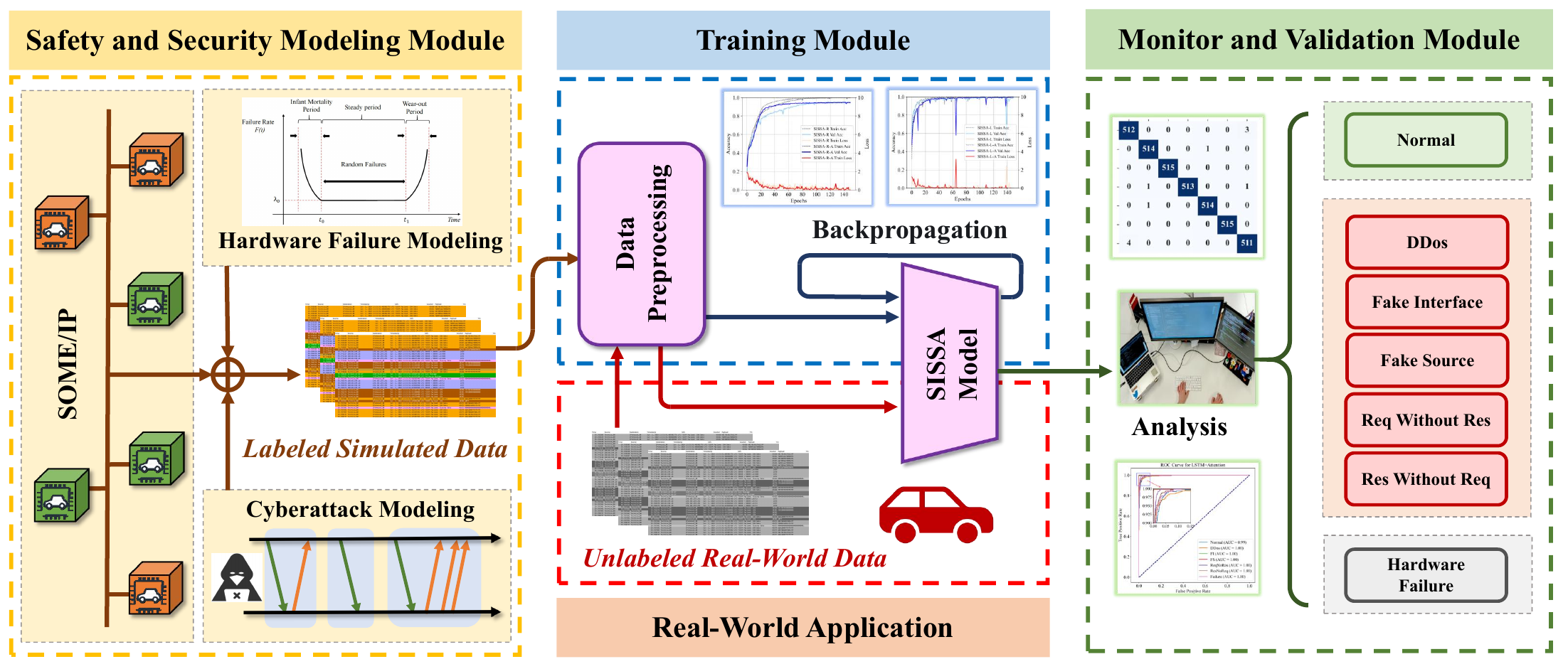}
\caption{Workflow of SISSA}
\label{fig:safety and security monitoring model}
\end{figure*}

%\subsection{Overview of the proposed scheme}
This Section proposes a novel safety and security monitoring framework, as illustrated in Figure \ref{fig:safety and security monitoring model}. ECUs communicate with each other via the SOME/IP protocol in Ethernet, which can be susceptible to functional failures and cyberattacks. We firstly model scenarios involving functional failure and cyberattck in the communication of ECUs through SOME/IP. Subsequently, training module is developed using unlabeled real-world data, employing deep learning models to extract characteristics from SOME/IP sessions among ECUs. We incorporate a residual self-attention block to accelerates the model's convergence and enhances detection accuracy. Finally, the targeted ECU undergoes analysis to determine whether it is functioning normally, experiencing functional failure, or under attack. In the following sections, we will detail the contents. 

\subsection{Safety and security modeling module }
\subsubsection{Modeling functional failure scenarios}
%\textbf{Malfunction Model}

ISO 26262 defines functional safety as the mitigation of risks arising from wear and tear resulting from abnormal operation of electrical and electronic systems. The standard outlines two primary types of failures in electrical/electronic systems: systematic failure and random hardware failure \cite{ISO-26262}. Systematic failure, associated with a specific cause, can only be addressed through modifications to the design, manufacturing process, operational procedures, documentation, or other related factors. According to ISO 26262, quantifying systematic failures is typically challenging using probabilistic methods. Developers achieve comprehensive system fault diagnosis with in-vehicle online monitoring, detection, and alerts to internal faults, coupled with off-vehicle offline diagnostics extracting fault information externally through  Unified Diagnostic Services requests for investigation and maintenance \cite{autosar_diagnostic}.

While random hardware failure refers to unexpected failures during the hardware element's lifecycle, following a probability distribution  \cite{ISO-26262}. ISO 26262 introduces the Probabilistic Metric for Hardware Failure (PMHF) in its 5th and 10th parts. PMHF is employed to assess the Automotive Safety Integrity Level, a measure of the probability of random hardware failures. It serves as a crucial target value in the hardware design of ECUs. In this work, we concentrate on the probabilistic metric for random hardware failures. Differing from in-vehicle system fault diagnosis, our objective is to characterize the variations of random hardware failures throughout the lifecycle, aiming to identify potential hardware malfunctions as comprehensively as possible. 
%\textbf{Malfunction Model}
\begin{figure}[!t]
	\centering
	\includegraphics[width=3.5in]{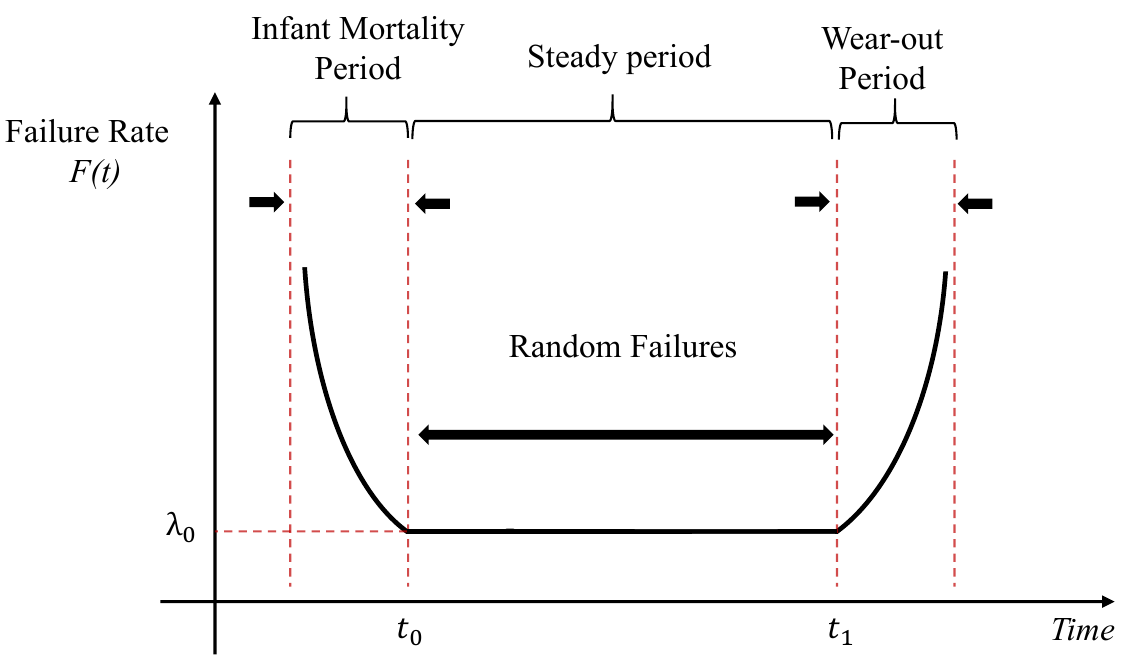}
	\caption{The bathtub curve of random hardware failures evolution}
	\label{fig: bathtub}
\end{figure}

We model random hardware failures based on the observation: a significant correlation between automotive product failure rates and their usage time or driving mileage throughout their lifecycle. The failure rate curve often exhibits a bathtub shape, featuring higher failure rates at both ends and a lower rate in the middle \cite{prasanth2017demystifying}. This pattern can be categorized into three stages as depicted in Figure \ref{fig: bathtub}.
The initial stage is the infant mortality period, marked by a high initial failure rate that decreases rapidly with accumulated operational time. Subsequently, the system enters the random failures or steady period, characterized by an extended duration with a low and stable failure rate. In this phase, persistent failures occur, termed as the random failure stage, which cannot be entirely eliminated irrespective of design quality. The final stage is the wear-out period, where the failure rate experiences a rapid increase over time due to factors such as wear, fatigue, aging, and consumption.

\begin{figure}[!t]
	\centering
	\includegraphics[width=3.3in]{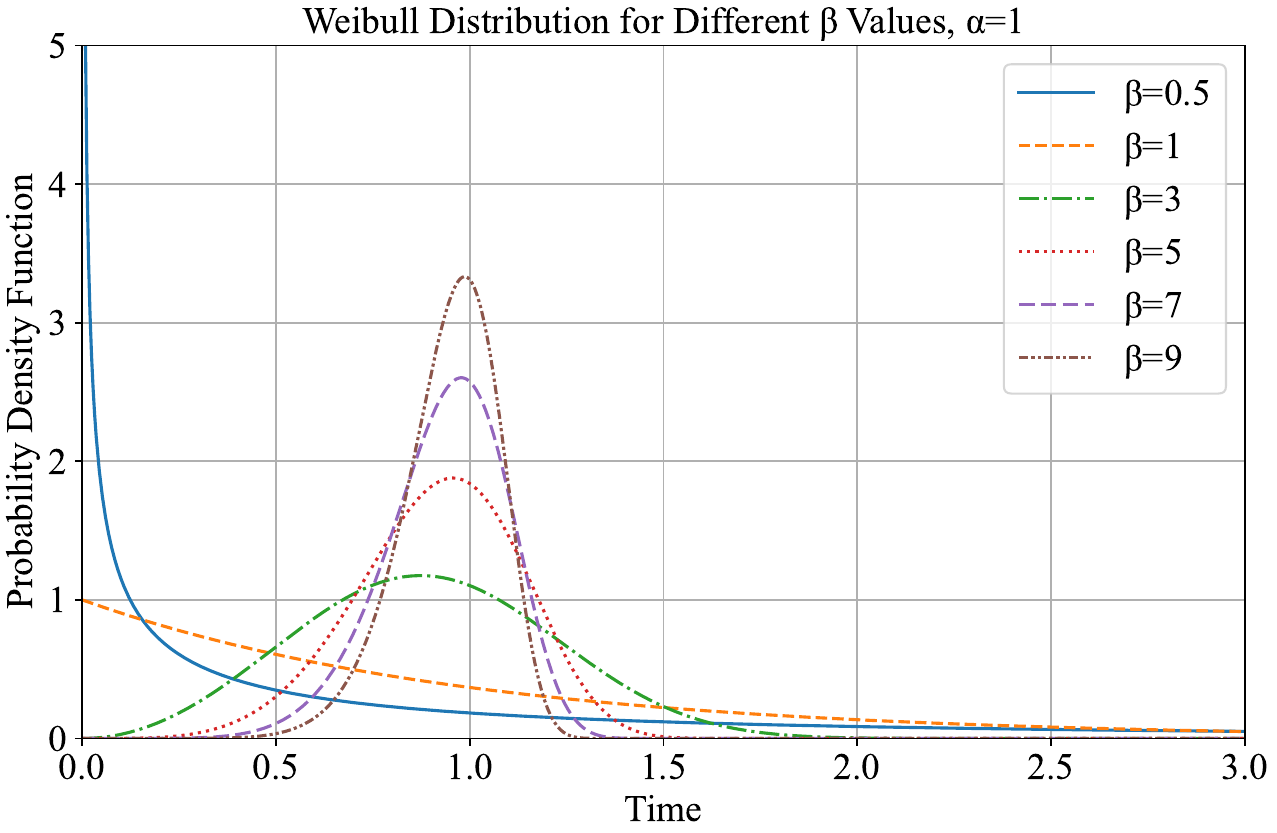}
	\caption{The Weibull distribution under different shape parameter}
	\label{fig: weibull}
\end{figure}

For the purpose of calculations, ISO 26262 posits that each component within the system exhibits a consistent failure rate throughout its operational lifespan. This is corresponding to the  steady period of random hardware failure evolution depicted in Figure \ref{fig: bathtub}. In this paper, to broaden the scope of random hardware failure analysis, we assume the random hardware failures follow a non-exponential trend based on the observations justified in Figure \ref{fig: bathtub}. The failure pattern can be described by the Weibull distribution \cite{kleyner2018calculating}, which are commonly used in reliability analysis. The probability density function of Weibull is:
\begin{IEEEeqnarray}{c}
	\label{deqn_ex1}
f(t) = \frac{\beta}{\alpha} \left(\frac{t}{\alpha}\right)^{\beta-1} e^{-\left(\frac{t}{\alpha}\right)^\beta}
\end{IEEEeqnarray}
where $\beta$ is the shape parameter, which determines the probability of the shape of the precision curve. $\alpha$ is the scale parameter, indicating the scaling of the function.

As depicted in Figure \ref{fig: weibull}, adjusting the shape parameter $\beta$ enables the simulation of the three stages corresponding to the bathtub effect throughout the entire product lifecycle. These stages include early failures, random failures, and wear-out failures.

\begin{itemize}
	\item When $\beta$ \textless 1, the failure rate exhibits a decreasing function of time $t$, suitable for modeling early failures in automobiles. 
	\item When $\beta$ = 1, the failure rate approximates a constant, applicable for modeling random failures during the normal usage period of automobiles. 
	\item When $\beta$ \textgreater 1, the failure rate demonstrates an increasing function of $t$, suitable for modeling wear-out failures in automobiles. This primarily manifests as failure issues occurring in products/components entering a period of wear and aging after prolonged operational time.
\end{itemize}

%It’s noted that calculated probability of random hardware failures can be compared with the random hardware failure target values shown in Table 1, to determine whether the system meets the required ASIL level.
\begin{figure*}[!t]
	\centering
	\includegraphics[width=7.1in]{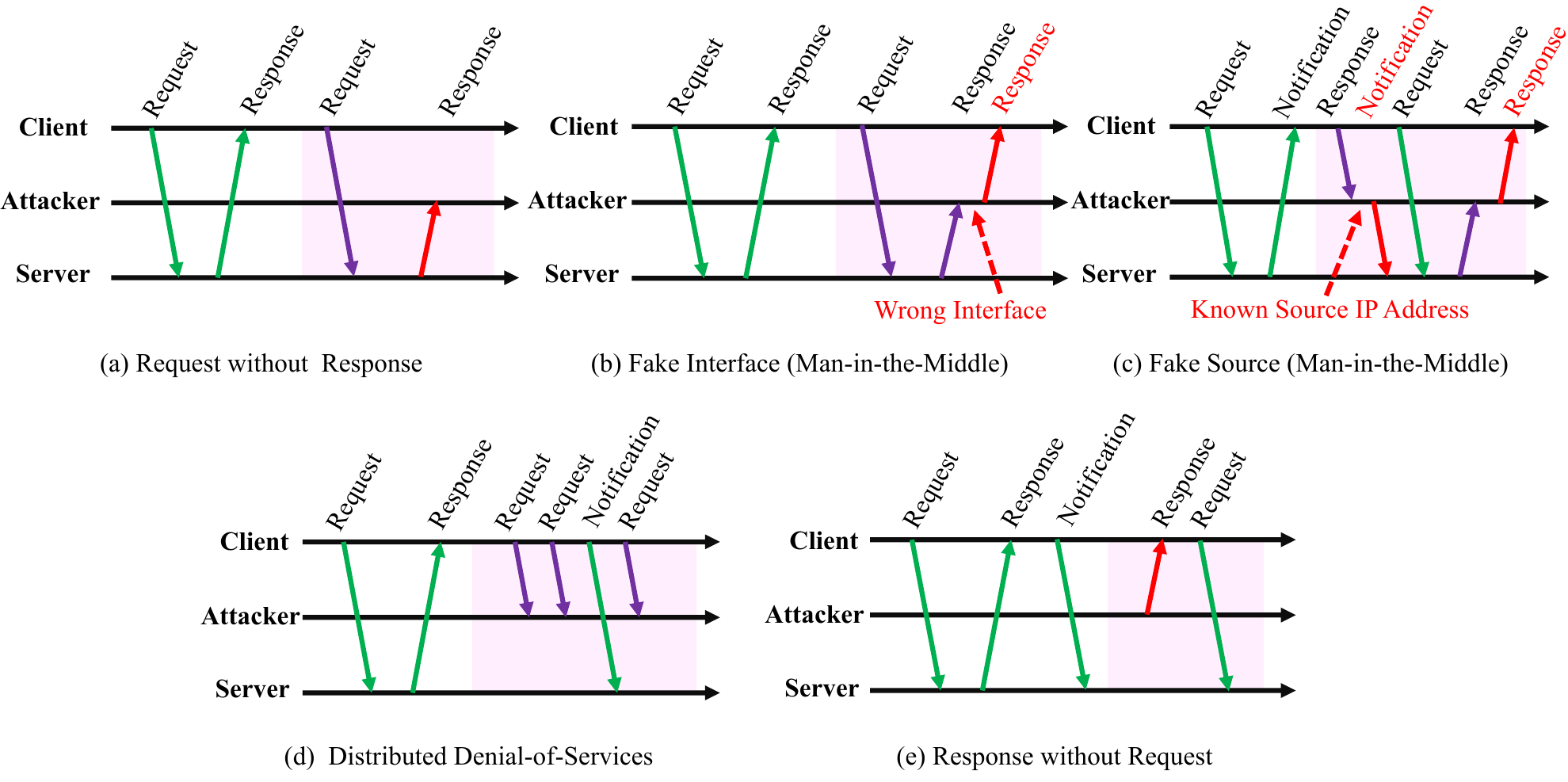}
	\caption{The vulnerabilities of SOME/IP and construct the attack scenario, including (a) (e) abnormal communication process, (d) Distributed Denial-of-Services attack, (b) (c) Man-in-the-Middle attack}
	\label{fig:attacks}
\end{figure*}

\subsubsection{Cyberattack scenarios}
In this work, we assume that the attacker gains access to the in-vehicle Ethernet network. Potential entry points may be discovered through software compromise or by connecting a physical device to the in-vehicle network. The attacker has the capability to capture broadcast and multicast transmissions in a switched network, implying that SOME/IP packets are accessible to them. We exploit the vulnerabilities of SOME/IP and construct the attack scenarios as illustrated in Figure \ref{fig:attacks}, encompassing Man-in-the-Middle (MITM) attacks, Distributed Denial-of-Service (DDos) attacks, and abnormal communication processes.

\begin{itemize}
	\item \textbf{Request without  Response}: The attacker interrupts response packets. In this case, the client cannot receives a certain amount of response packets, leading to abnormal client behavior.
	\item \textbf{Fake Interface}: The attacker identifies response packets and injects nearly identical packets, with the only distinction being an incorrect interface.
	\item \textbf{Fake source}: The attacker replicates a chosen packet, injects it into the communication, and substitutes the original source IP address with another valid one. The new address is randomly selected from existing legal addresses, effectively deceiving the address checking.
	\item \textbf{DDos}: The adversary eavesdrops on SOME/IP communication and disrupts client-to-server request packets at higher rate than normal. As a result, the client fails to receive a certain number of response packets from the server. 
	\item \textbf{Response without Request}: The attacker sends response packets to a client in the name of a server before the client sends request to the server. For instance, attacker injects a response packet from a server to client when the client has just sent a notification packet to the server and is about to send a request. 
\end{itemize}

\subsection{Training module}
%This section examines the SOME/IP packet exchange in the Ethernet to determine whether an Electronic Control Unit (ECU) is under attack, experiencing functional failure, or operating normally. CNN and RNN are adopted in this paper as they show remarkable classification performance \cite{liu2019cnn}. In the realm of neural networks, attention serves as a technique that emulates cognitive focus. Implementing attention mechanisms within deep neural networks enables rapid identification of key elements in the target amidst a vast amount of information, facilitating the elimination of irrelevant data and enhancing the network's efficiency in task execution.   

\begin{figure*}[!t]
	\centering
	\includegraphics[width=7.1in]{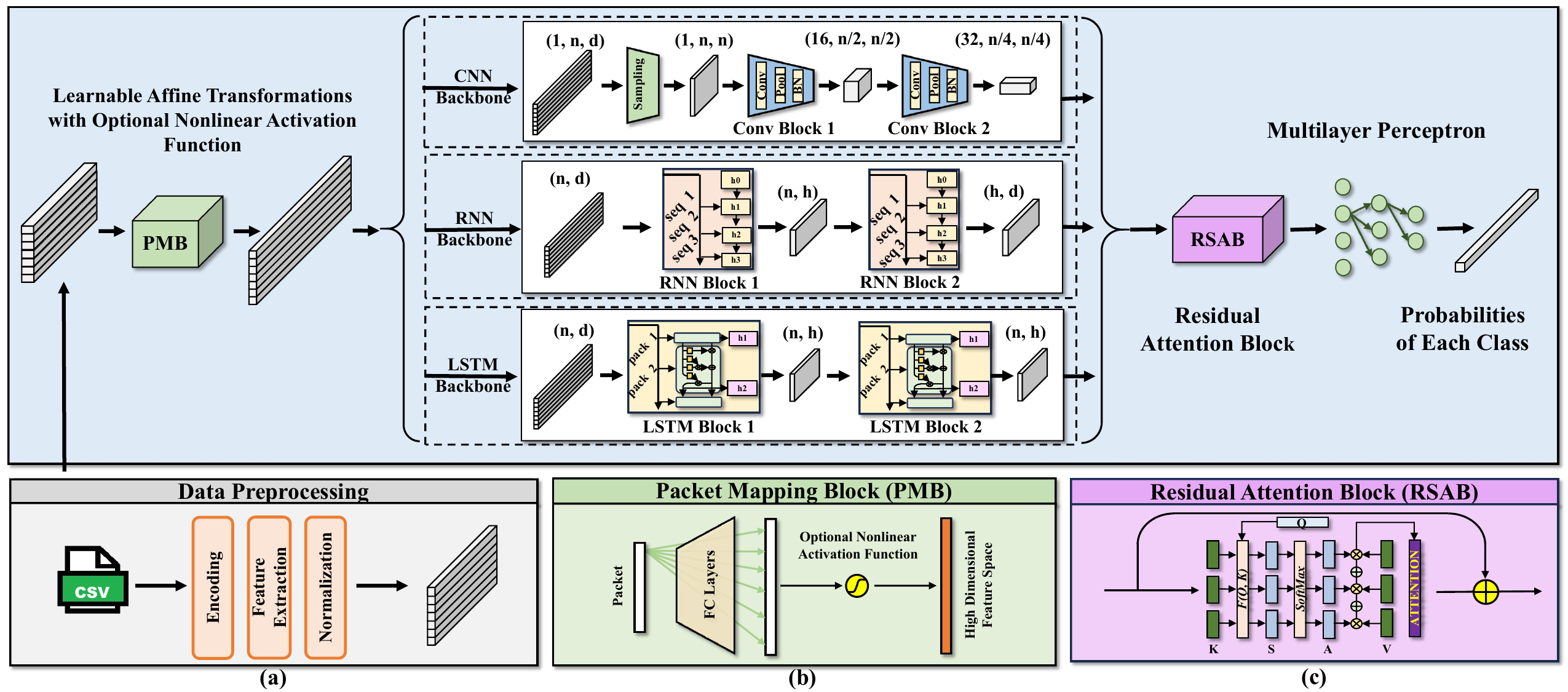}
	\caption{(Top) The training process. (Bottom) Details of (a) Data preprocessing, (b) Packet Mapping Block, (c) Residual Attention Block.}
	\label{fig:data learning}
\end{figure*}
\subsubsection{ Data preprocessing }
Figure \ref{fig:data learning} depicts the process of data training, designed for the analysis of $n$ time-stamped messages within a detection window to perform multiple classification tasks. Within our SOME/IP data described in Section \ref{some/ip protocol}, each message is characterized by 10 specific message fields plus a temporal attribute. Critical fields, such as 'Message Type' and 'Error Code', undergo OneHot encoding to convert them into binary vectors, a process essential for the discrete categorization inherent in these fields. This enables the network to more effectively identify the interrelations between disparate values. Notably, OneHot encoding is selectively employed for particular fields only, which leads us to design the Packet Mapping Block (PMB) to extract the feature of other continuous-values.

\subsubsection{Packet mapping }
After the initial data preprocessing step, the PMB is applied. The PMB utilizes learnable affine transformations $A(m) = Wm + b$, where $W$ and $b$ denote the weight matrix and bias vector respectively, to incrementally extend the dimensionality of each message to a $d$-dimensional feature space. As a result, the input window is reshaped to a dimensionality of $n$$\times$$d$.

The above two step is a basic progress to preprocess a window of Some/IP packets. Next, depending on the choice of backbone, the data would be sent to different structures for futher feather extraction and classification.

%Subsequent to the PMB, the data traverses through the network's backbone structures which vary depending on the specific SISSA network configuration. This path may include the integration of an optional Residual Self-Attention Block (RSAB),
%\begin{IEEEeqnarray}{c}
%	\label{deqn_ex1}
%	RSAB(x) = x + \text{Attention}(x)
%\end{IEEEeqnarray}
% which incorporates a skip connection around the attention mechanism to enhance learning of dependencies. The process is completed with a Multi-Layer Perceptron (MLP):
%\begin{IEEEeqnarray}{c}
%	\label{deqn_ex2}
%     MLP(x) = \sigma(W^{(L)} (\ldots \sigma(W^{(1)} x + b^{(1)}) \ldots) + b^{(L)})
%\end{IEEEeqnarray}
%
%where $\sigma$ denotes a nonlinear activation function, and $W(i)$ and $b(i)$ represent the weights and biases of the ith layer. The MLP outputs the probability distribution over the various classes. we will detail the 
%design of network's backbone and RSAB in tbe following part.

\subsubsection{CNN backbone}
Convolutional Neural Networks (CNNs) have been investigated for their efficacy in intrusion detection \citenum{zhao2022semi}. It is hypothesized that when message windows are sampled across continuous timestamps, they will display discernible image patterns that differentiate normal behavior from anomalies. Such a premise paves the way for leveraging CNN-based architectures to discern and categorize intrusions.

In the context of our proposed models, SISSA-C and SISSA-C-A, we employ a CNN backbone as the principal mechanism for feature extraction and pattern recognition. The detailed schematic of this CNN backbone is elucidated in Figure \ref{fig:data learning}. Input data, represented in a $n$$\times$$d$ matrix, undergoes an initial transformation to a 3-dimensional tensor with dimensions $1$$\times$$n$$\times$$d$, where the singleton dimension symbolizes a single-channel grayscale image. Subsequently, a bespoke sampling algorithm reformats the data into a  $1$$\times$$n$$\times$$n$ image.

This processed image is then propagated through a sequence of convolutional blocks. Each block comprises a convolutional layer (Conv), followed by a Max Pooling layer (MaxPool), and a Batch Normalization layer (BN). The operations within these blocks can be formally described as follows:
for a given convolutional block $B_i$,
the transformation $T$ applied to the input $I$ is:
\begin{IEEEeqnarray}{c}
	\label{deqn_ex1}
	T(I) = \text{BN}(\text{MaxPool}(\text{Conv}(I)))
\end{IEEEeqnarray}

To encapsulate inter-channel relationships and the intricate feature interplay, the channels corresponding to each pixel are subject to a non-linear mapping. This mapping, designed to extract the nuanced associations between diverse channels, further refines the model's intrusion detection capabilities.
The non-linear mapping $M$ between the channels for a pixel $p$ is expressed as:
\begin{IEEEeqnarray}{c}
	\label{deqn_ex1}
	M(p) = f(p_1, p_2, \ldots, p_c)
\end{IEEEeqnarray}
where $f$ represents the non-linear function, and $p_1, p_2, \dots, p_c$ are the channel values of pixel $p$ across $c$ channels.

The amalgamation of these computational layers and the non-linear mapping engenders a robust framework capable of identifying anomalous patterns indicative of cyber attack and random hardware failure. 

\subsubsection{RNN backbone}
Although CNNs have been a conventional choice for intrusion detection, their efficacy in interpreting sequential data remains limited. Different from that, sequence-based neural networks could potentially offer superior generalization capabilities, particularly in the context of processing continuous message streams \cite{chen2021flag}.

Figure \ref{fig:data learning} delineates the architecture of our proposed sequence-based models, SISSA-R and SISSA-R-A. These models integrate Recurrent Neural Network (RNN) blocks, designed to capture temporal dependencies and extract timing features from sequences of network packets. The RNN blocks are structured to process an input window comprising $n$ packets, with each packet represented as a vector of dimension $d$.

In the context of SOME/IP communication, encompassing both attack and fault scenarios, we carefully adhere to the temporal accuracy of packet transmission. This temporal aspect may elude the feature detection capacities of purely image-based pattern recognition systems. Therefore, we introduce a multi-layered RNN architecture to address this challenge. The architecture functions as follows.

For each RNN block $B_i$, the transformation $T$ across a sequence of length $n$ is defined recursively by:
\begin{IEEEeqnarray}{c}
	\label{deqn_ex1}
	h_t = f(h_{t-1}, x_t; \theta)
\end{IEEEeqnarray}
where $h_t$ is the hidden state at time $t$, $x_t$ is the input at time $t$, and $\theta$ represents the parameters of the RNN.

The schematic progression can be summarized as:
\begin{IEEEeqnarray}{c}
	\label{deqn_ex2}
	(n, d) \xrightarrow{\text{RNN Block 1}} (n, h) \xrightarrow{\text{RNN Block 2}} (h, d)
\end{IEEEeqnarray}
Each RNN block captures and propagates temporal information through its hidden states, thereby facilitating an intricate understanding of the temporal patterns characteristic of intrusion scenarios. By harnessing this architecture, SISSA-R and SISSA-R-A aim to provide a nuanced analysis of network traffic, potentially outperforming traditional CNN-based approaches in detecting sophisticated cyber attacks.

\subsubsection{LSTM backbone}
 The  Long Short-Term Memory (LSTM) is capable of  capturing long-range dependencies within a sequence, making it eminently suitable for scenarios where distant messages within a window may exert a significant impact \cite{gao2020omni}. In the field of intrusion detection, the LSTM's architecture is adept at disentangling and learning from the intricate temporal patterns that characterize different attack vectors and fault conditions. The ability of LSTM to mitigate the vanishing gradient problem prevalent in traditional RNNs allows for more effective learning over extended sequences, which is paramount in the nuanced classification tasks presented by multiple attack and fault modes.

Figure \ref{fig:data learning} illustrates the LSTM backbone composed of multiple LSTM blocks. These blocks are embedding into SISSA-L and SISSA-L-A models. Each block processes an input sequence of \( n \) packets, with each packet characterized by a \( d \)-dimensional feature vector. The LSTM block is engineered to iteratively update its cell states and hidden states through a series of gated operations—input, output, and forget gates, thus selectively retaining or discarding information through the sequence.

Formally, the LSTM updates for each time step \( t \) in a given block \( B_i \) can be described by the following equations:
\begin{IEEEeqnarray}{c}
	\label{deqn_ex3}
	\begin{IEEEeqnarraybox}[][c]{rCl}
		f_t & = & \sigma(W_f \cdot [h_{t-1},x_t] + b_f) \\
		i_t & = & \sigma(W_i \cdot [h_{t-1}, x_t] + b_i) \\
		\widetilde{C}_t & = & \tanh(W_c \cdot [h_{t-1}, x_t] + b_C) \\
		C_t & = & f_t * C_{t-1} + i_t * \widetilde{C}_t \\
		o_t & = & \sigma(W_o \cdot [h_{t-1}, x_t] + b_o) \\
		h_t & = & o_t * \tanh(C_t)
	\end{IEEEeqnarraybox} \\
	\mbox{} \nonumber
\end{IEEEeqnarray}
where \( f_t \), \( i_t \), and \( o_t \) represent the forget gate, input gate, and output gate activations respectively, \( C_t \) and \( h_t \) are the cell state and hidden state, \( \sigma \) denotes the sigmoid function, and \( W \) and \( b \) are the weights and biases associated with each gate. 

The architectural schematic translates to a transformation pathway:

\begin{IEEEeqnarray}{c}
	\label{deqn_ex2}
	(n, d) \xrightarrow{\text{LSTM Block 1}} (n, h) \xrightarrow{\text{LSTM Block 2}} (n, h)
\end{IEEEeqnarray}
Each LSTM block's output, \( h_t \), encodes not only the information of the current packet but also the context of preceding packets. This makes it a valuable resource for identifying patterns that signal anomalies, even in inputs with intricate temporal dynamics. 

By harnessing the LSTM backbone within our proposed models, SISSA-R and SISSA-R-A, we aim to elevate the accuracy and generalization of monitoring system to adeptly handle the multifarious and temporally complex patterns observed in functional failure and attack.

\subsubsection{ Residual Self-Attention Block (RSAB)}

RSAB aims to enhance the network's ability to discern complex and subtle relationships within the data. This is crucial for differentiating patterns indicative of attacks, which are typically intentional and well-organized, from the stochastic nature of hardware failures, which might manifest as irregular and less predictable patterns.

Attacks can occur through coordinated sequences of messages that may seem legitimate when considered individually. The RSAB, with its intrinsic attention mechanism, is adept at capturing the sequential context across the entire window of messages, facilitating the identification of such coordinated anomalies. It achieves this by calculating attention weights, which essentially determine how much focus should be placed on each part of the input sequence when constructing a representation of a given message.

Mathematically, the attention mechanism within the RSAB can be expressed as:
\begin{IEEEeqnarray}{c}
	\label{eqn_attention}
	\text{Attention}(x) = \text{Softmax}\left(\frac{QK^T}{\sqrt{d_k}}\right)V
\end{IEEEeqnarray}
where \( Q \), \( K \), and \( V \) are the query, key, and value matrices derived from the input data \( x \), and \( d_k \) is the scaling factor corresponding to the dimension of the keys. This attention function allows the network to prioritize certain messages within a window, which is crucial when a single malicious packet could indicate a potential attack.

Conversely, in diagnosing random hardware failures, the RSAB aids in discerning non-obvious correlations across time-stamped messages. Hardware failures may not follow a recognizable pattern, and as such, the network must be sensitive to unusual deviations that could be indicative of such faults. The RSAB empowers the SISSA network to maintain a memory of past message features that, while seemingly unrelated, may collectively signify a hardware malfunction.

The inclusion of the residual connection, formulated as equation \ref{eqn_attention}, enables the preservation of the original features while incorporating the refined attention-driven representations. This aspect of the RSAB is particularly advantageous in mitigating the risk of over-smoothing features—a common challenge in deep networks, where the model may lose the ability to differentiate between classes due to excessive blending of feature representations.

\section{Evaluation}
\label{Evaluation}
\subsection{Experimental Settings}
\subsubsection{Dataset}
In our comprehensive open-source framework, SISSA, we delineate a holistic approach encompassing the generation of standard data, the simulation of various attacks and random hardware failures, alongside meticulous data preprocessing and network training methodologies.

The Some/IP Generator, a Python-based tool for generating Some/IP packets, is available on GitHub and has been extensively utilized in relevant intrusion detection \cite{SOMEIPGenerator}\cite{alkhatib2023said}. This generator employs Python's native multiprocessing components and multiple communication queues to effectively simulate the traffic of Some/IP packets among various ECUs within the same Local Area Network over continuous time. However, due to the prolonged lack of maintenance of the GitHub repository, the generator faces challenges in implementing modern, complex message behaviors. For instance, the packet generation process does not fully adhere to real-world timing constraints, rendering it inadequate for simulating attacks that are highly time-sensitive. Moreover, the generator's current design does not effectively represent the random hardware malfunctions in the communication queue.

Given these limitations, we propose enhancements to the Some/IP Generator. Our approach leverages its multi-ECU communication simulation capabilities to first generate standard Some/IP traffic packets. Subsequently, we aim to model in-vehicle functional safety and cyber security threats. This will be achieved through custom scripts that can generate specific attack message windows and simulate random hardware failure message windows, thereby providing a more robust and realistic testing environment for monitoring systems.

\begin{figure}[!t]
	\centering
	\includegraphics[width=3.5in]
	{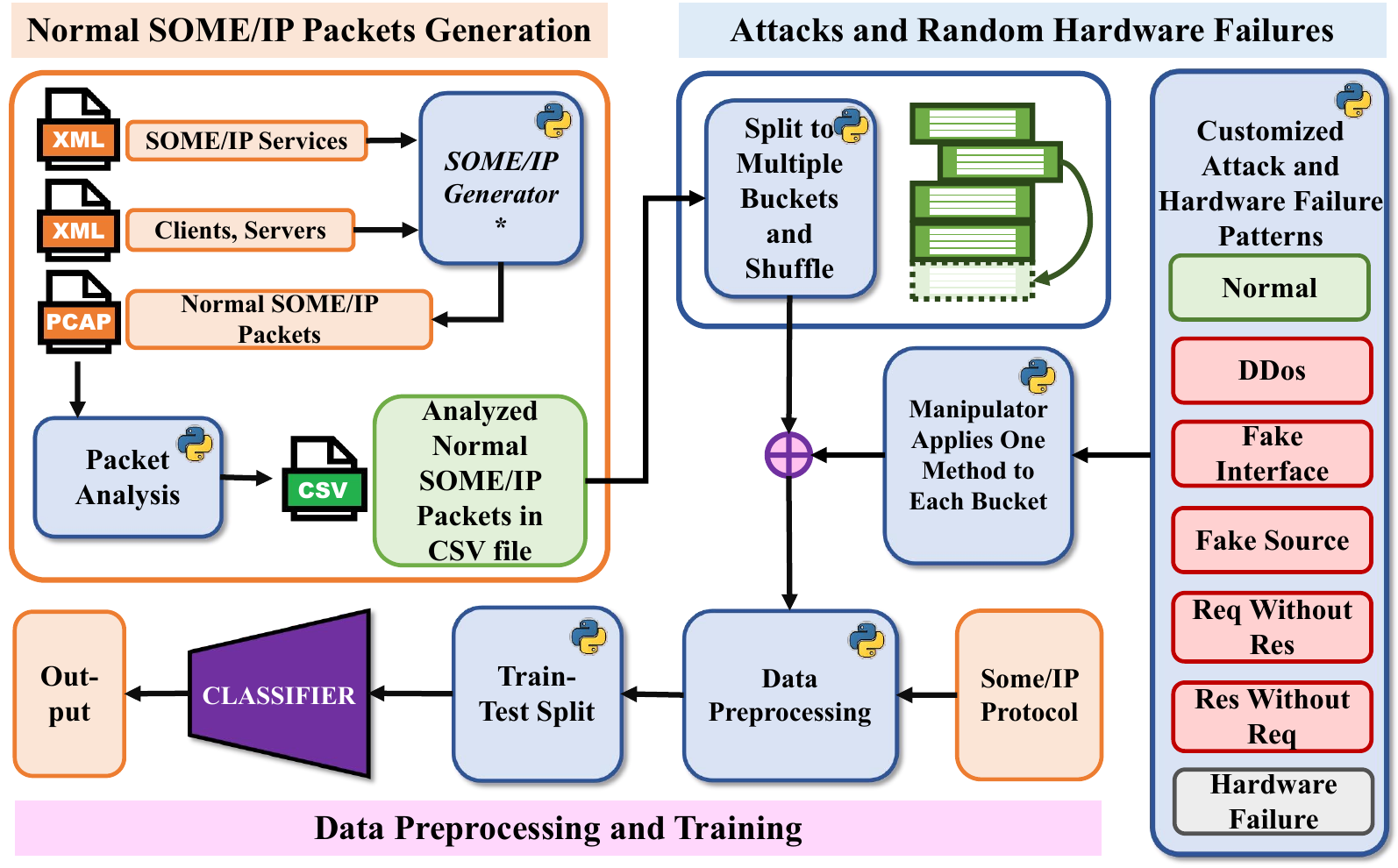}
	\caption{Dataset generation process for SOME/IP }
	\label{fig: Dataset  generation}
\end{figure}

Figure \ref{fig: Dataset  generation} shows the dateset generation process for SOME/IP. By utilizing the SOME/IP Generator, we initiate the process by generating a corpus of typical SOME/IP communication data.  This data is subsequently processed through our data segmentation module.  In this phase, the continuous stream of data is partitioned into distinct blocks. These blocks are subjected to simulated attack or failure scenarios after shuffling.  Subsequently, each block is further subdivided into multiple, equal-length windows, allocated to respective buckets based on the classification task's category.  In our case, this results in seven buckets, each containing diverse Windows subjected to identical attack simulations.  To enhance the realism and variability of these simulations, we introduce adjustable parameters, ensuring that each attack instance retains a unique characteristic.

\begin{table}
	\centering
	\caption{Training and testing dataset classes}
	\label{table:Dateset}
	\begin{tabular}{c|c|c}
		\hline
		\hline
		Class & Training dateset & Validation dateset \\
		\hline
		Normal & 2058*128 & 515*128 \\
		DDos & 2058*128 & 515*128 \\
		Fake Interface & 2058*128 & 515*128 \\
		Fake source & 2058*128 & 515*128 \\
		Req without Res & 2058*128 & 515*128 \\
		Res without Req & 2058*128 & 515*128 \\
		Hardware Failure & 2058*128 & 515*128 \\
		Total & 2058*128*7 & 515*128*7 \\
		\hline
		\hline
	\end{tabular}
	
    \begin{tablenotes}
	\footnotesize
	%\item[*] this is the ....  %此处加入注释*信息
	%\item[**] my website is ... %此处加入注释**信息
	\item Req: Request; Res: Response.

\end{tablenotes}
\end{table}

Table \ref{table:Dateset} shows the details of our training set and validation set window data. A critical aspect of our approach is the strict equalization of windows across different categories, a measure implemented to preclude any potential training bias due to data imbalances at the simulation stage.  Additionally, the entire process of attack and failure simulation is optimized through multi-process execution, significantly enhancing the speed of data generation and experimental efficiency.

Upon the generation of bucketed data, we engage in a series of preprocessing and feature engineering steps.  The window data within each bucket is then randomized and split into training and validation sets in an 8:2 ratio.  Once segmented into these sets and saved as .npy files, the data is primed for neural network training.

It is worth mentioning that the SISSA framework is of high  extensibility. Being entirely open-source and transparent, it allows researchers the flexibility to modify modules to suit specific objectives.  Although our focus is on SOME/IP packet detection and analysis, the framework is adaptable to other data types, such as CAN.  Researchers can introduce custom attack and failure modes, and reconfigure the network structure for retraining the classifier.  Moreover, the management of all parameters is facilitated through several YAML files, obviating the need for code modification within Python scripts once the model structure is established.  This feature significantly enhances the usability and accessibility of the SISSA framework for a wide array of research applications.

\begin{figure*}[!t]
	\centering
	\includegraphics[width=\linewidth]{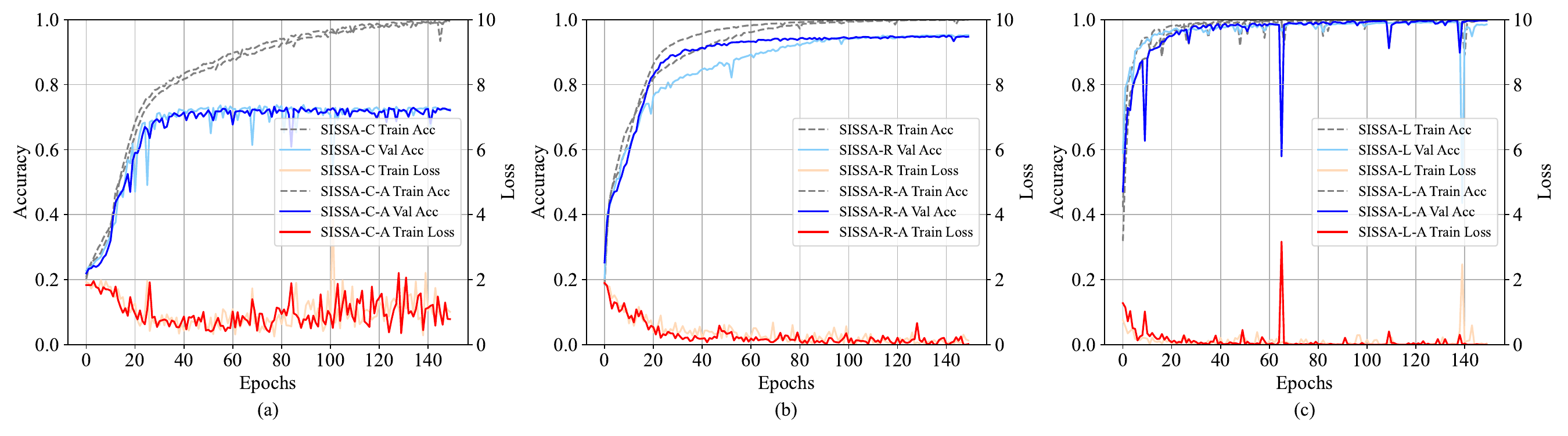}
	\caption{Train accuracy and loss of each SISSA models. (a) indicates the CNN  based models, (b) indicates the RNN  based models, (c) indicates the LSTM  based models}
	\label{fig: Train loss and acc}
\end{figure*}

\subsubsection{Metrics}
Classification task contains four possible outcomes.  True Positive (TP) model accurately identifies positives. True Negative (TN) model correctly identifies negatives. False Positive (FP) models wrongly labels instances as positives. False Negative (FN) models fails to identify positives. Based on TP, TN, FP and FN, we evaluate SISSA’s performance using the following metrics:
\begin{itemize}
	\item \textbf{Accuracy} measures the overall correctness of a model by assessing the ratio of correctly predicted instances to the total instances, providing a general performance indicator ($\frac{TP+TN}{TP+TN+FP+FN}$). 
	\item \textbf{Recall} quantifies the model's ability to correctly identify all relevant instances among the actual positives, emphasizing the avoidance of false negatives ($\frac{TP}{TP+FN}$). 
	\item \textbf{Precision} gauges the model's accuracy in labeling instances as positive, emphasizing the avoidance of false positives and assessing the relevance of the identified positives ($\frac{TP}{TP+FP}$). 
	\item \textbf{F1-score} is the harmonic mean of precision and recall, offering a balanced measure that considers both false positives and false negatives, suitable for evaluating models in scenarios with imbalanced class distribution ($\frac{2\times Precision \times Recall}{Precision + Recall}$). 

	\item \textbf{ROC} Curve assesses model performance in neural networks. It plots False Positive Rate and True Positive Rate for different discrimination thresholds. The ideal ROC curve should be closer to the upper-left corner, indicating better performance. 
	\item \textbf{AUC} Scores quantifies the area under the right side of the ROC curve, providing a visual measure of the classifier's predictive performance.  In cases where ROC curves for different classes intersect, the AUC values serve as a quantitative criterion to determine which class exhibits superior predictive performance.

\end{itemize}

\begin{figure*}[!t]
	\centering
	\includegraphics[width=5.85in]{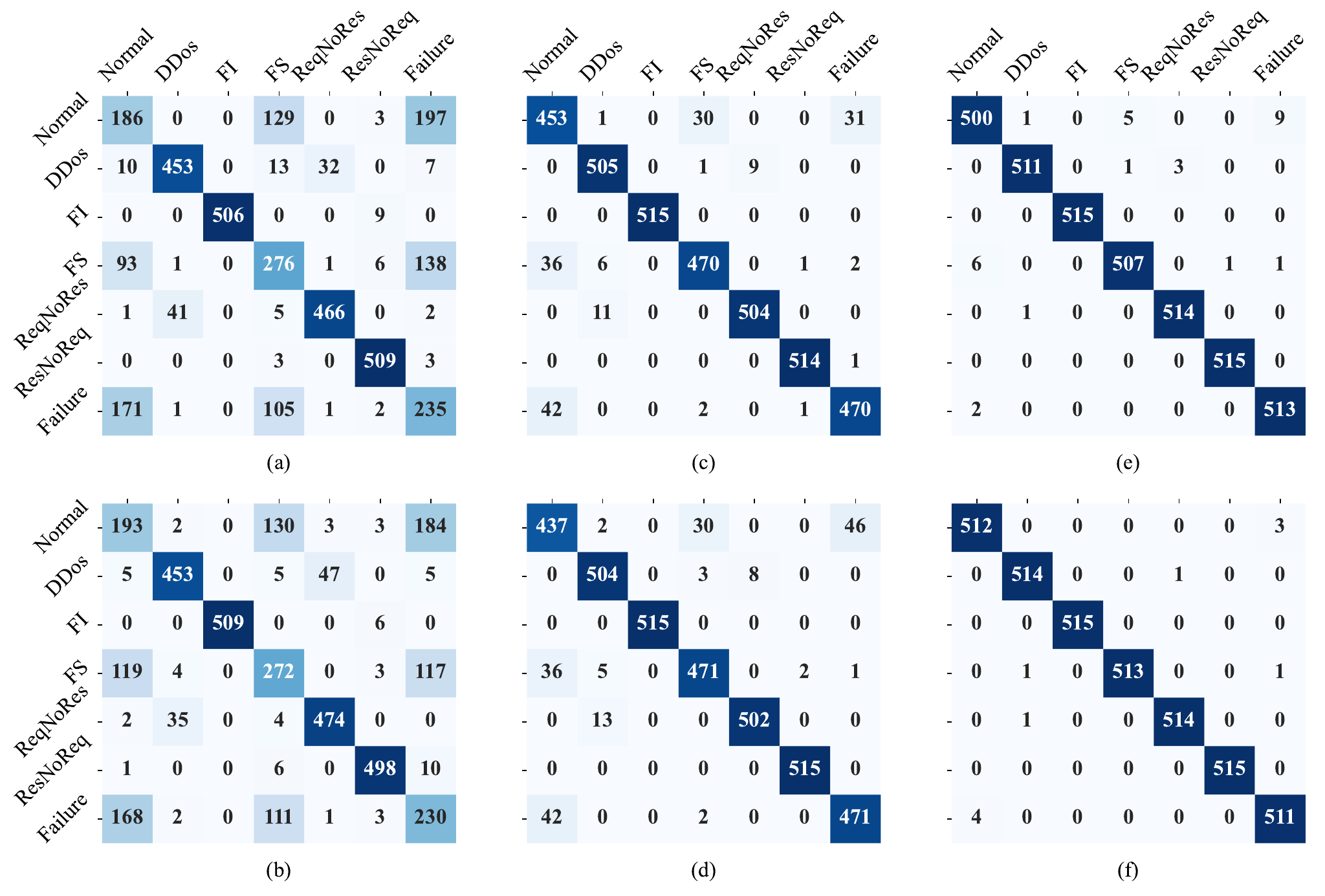}
	\caption{Confusion matrix for six different models on testing dataset. (a) and (b) indicate the CNN  based models, (c) and (d) indicate the RNN  based models, (e) and (f) indicate the LSTM  based models}
	\label{fig: Confusion}
\end{figure*}

\begin{figure*}[!t]
	\centering
	\includegraphics[width=\linewidth]{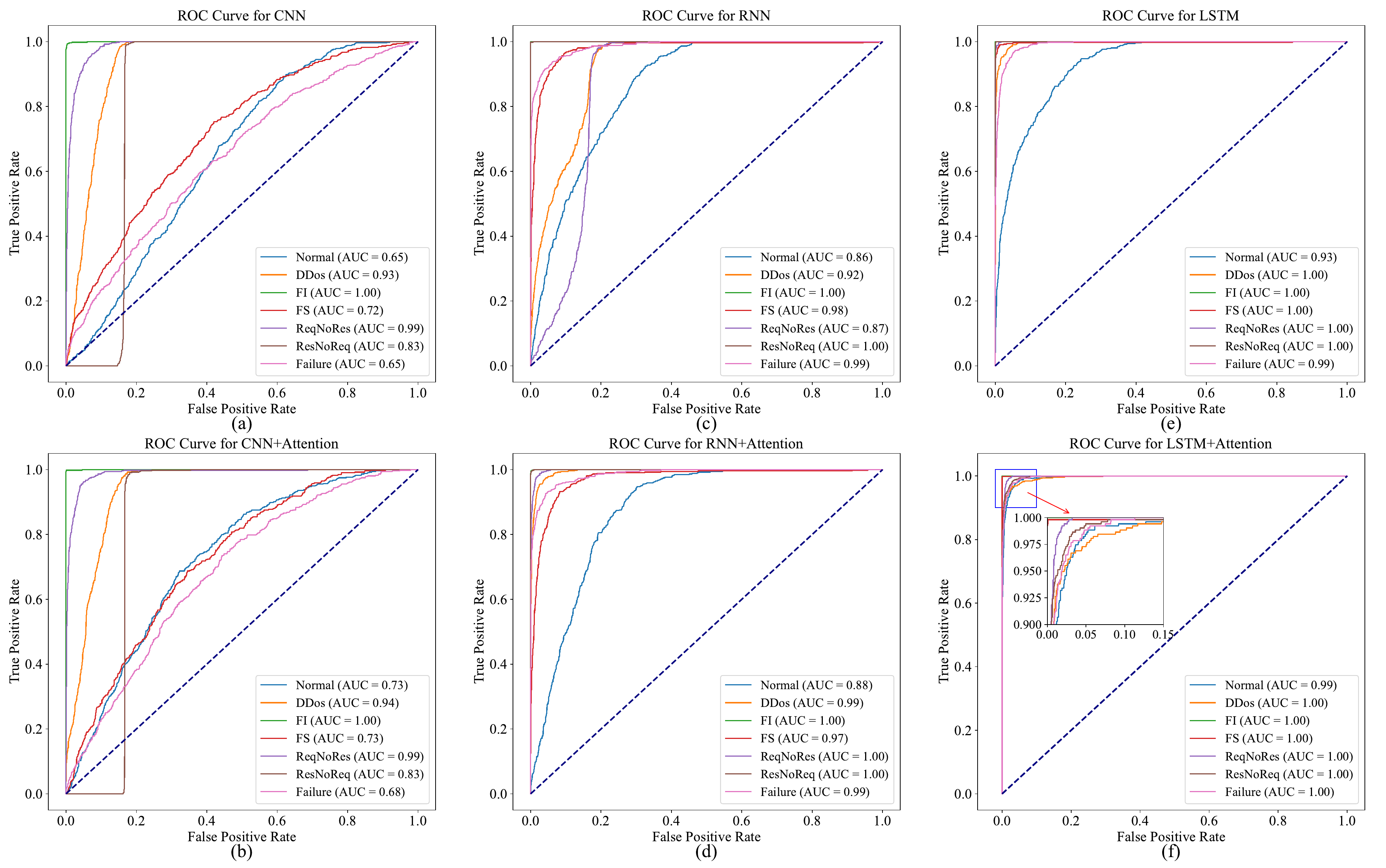}
	\caption{The ROC curves and AUC values for six models with different backbones across seven classification tasks. (a) and (b) indicate the CNN  based models, (c) and (d) indicate the RNN  based models, (e) and (f) indicate the LSTM  based models}
	\label{fig: ROC-AUC}
\end{figure*}

 \begin{figure*}[!t]
	\centering
	\includegraphics[width=5.8in]{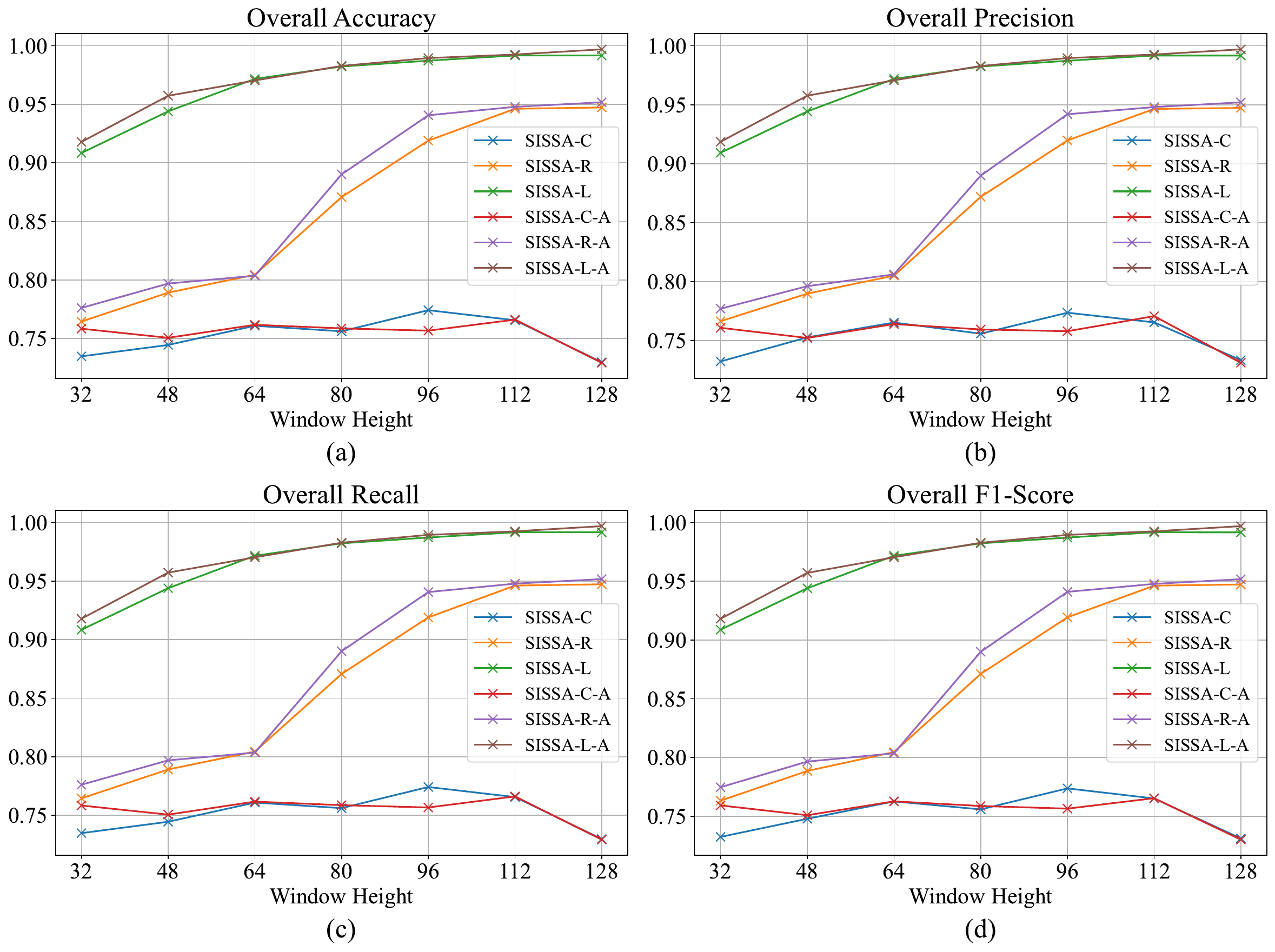}
	\caption{(a) Overal accuracy, (b) precision, (c) recall and (d) F1-score of each class on testing dataset under different widow size}
	\label{fig: Wsize}
\end{figure*}

\subsection{Experimental performance}
\subsubsection{Performance in training}

Figure \ref{fig: Train loss and acc} illustrates the training performance of six distinct SISSA models. These models were trained for a maximum of 150 epochs, with hyperparameters finely tuned within this constraint to optimize training efficacy. Of the six variants based on varying architectures, the SISSA-L-A model exhibited superior performance, achieving the highest validation accuracy of 99.7\% across seven classification tasks. Conversely, the CNN-based model demonstrated limited generalization capacity, though it attained a notable training accuracy of 99.5\%, translating to a lower validation accuracy of 72\%.

In scenarios Figure \ref{fig: Train loss and acc} (a) and (b), it is evident that the incorporation of an RSA layer at the network's terminus not only stabilizes the training process but also significantly enhances validation accuracy. Furthermore, scenario Figure \ref{fig: Train loss and acc} (c) reveals that the SISSA-L-A model, in comparison to its SISSA-L counterpart, exhibits improved generalization capabilities, particularly under extended training durations.

The stagnation in validation accuracy observed in the SISSA-C and SISSA-C-A models is attributed to the temporal nature of the attack and random failure modes delineated in our study. These temporal characteristics are inadequately captured by convolutional image classification neural networks, rendering them suboptimal for intrusion and fault detection tasks that exhibit specific characteristics.

The SISSA-L-A model stands out as our recommended solution for achieving the highest validation accuracy in seven distinct classification scenarios. Owing to its integration of LSTM and RSA modules, the SISSA-L-A is adept at extracting correlations between packet features across different timestamps in a window sequence. This capability enables it to effectively identify characteristics of SOME/IP traffic during vehicular attacks or random hardware failures.

\subsubsection{Performance in monitoring}

Figure \ref{fig: Confusion} presents the confusion matrices for each of the six evaluated models. An analysis of these confusion matrices reveals that both the SISSA-C and SISSA-C-A exhibit lower detection rates and higher false positive rates, particularly in scenarios involving DDoS attacks, Fake Source attacks, and random hardware failures. This diminished performance is attributed to the temporal dynamics inherent in these attack types, a feature that proves challenging for CNN-based models to effectively capture and analyze.

Conversely, SISSA-L-A demonstrates remarkable efficacy, as evidenced by the near absence of detection errors and false positives in Figure \ref{fig: Confusion}. This observation underscores the superior performance of the model, particularly in its ability to accurately identify and classify attack patterns and hardware malfunction. The results suggest that SISSA-L-A's architecture, which likely leverages advanced temporal feature extraction capabilities, is better suited for handling the complexities associated with these specific types of attacks.

\begin{table}
	\renewcommand{\arraystretch}{0.9}
	\centering
%	\captionsetup{font=Large} 
	\caption{The comparison of the  accuracy, precision, Recall and F1-score of each SISSA model}
	\label{table: comparison of each SISSA }
	\fontsize{8}{12}\selectfont
	\begin{tabular}{m{1.2cm} | >{\centering\arraybackslash}m{1cm} | >{\centering\arraybackslash}m{0.9cm} | >{\centering\arraybackslash}m{0.9cm} | >{\centering\arraybackslash}m{0.9cm} | >{\centering\arraybackslash}m{0.9cm}}
		\hline
		\hline
		\multicolumn{1}{c|}{Model} & \multicolumn{1}{c|}{Class} & \multicolumn{1}{c|}{Acc} & \multicolumn{1}{c|}{Pre} & \multicolumn{1}{c|}{Recall} & \multicolumn{1}{c}{F1} \\
		\hline
		\multirow{3}{1.2cm}{\centering SISSA-C} &  Normal & 0.832 & 0.403 & 0.361 & 0.381 \\
		\cline{2-6}
		&  Attack & 0.863 & 0.906 & 0.901 & 0.904 \\
		\cline{2-6}
		&  Failure & 0.826 & 0.404 & 0.456 & 0.428 \\
		\hline
		\multirow{3}{1.2cm}{\centering SISSA-C-A} &  Normal & 0.829 & 0.395 & 0.375 & 0.385 \\
		\cline{2-6}
		&  Attack & 0.857 & 0.901 & 0.899 & 0.900 \\
		\cline{2-6}
		&  Failure & 0.833 & 0.421 & 0.447 & 0.434 \\
		\hline
		\multirow{3}{1.2cm}{\centering SISSA-R} &  Normal & 0.961 & 0.853 & 0.880 & 0.866 \\
		\cline{2-6}
		&  Attack & 0.980 & 0.987 & 0.985 & 0.986 \\
		\cline{2-6}
		&  Failure & 0.978 & 0.933 & 0.913 & 0.922 \\
		\hline
		\multirow{3}{1.2cm}{\centering SISSA-R-A} &  Normal & 0.957 & 0.849 & 0.849 & 0.849 \\
		\cline{2-6}
		&  Attack & 0.980 & 0.987 & 0.986 & 0.986 \\
		\cline{2-6}
		&  Failure & 0.975 & 0.909 & 0.915 & 0.912 \\
		\hline
		\multirow{3}{1.2cm}{\centering SISSA-L} &  Normal & 0.994 & 0.984 & 0.971 & 0.978 \\
		\cline{2-6}
		&  Attack & 0.996 & 0.998 & 0.997 & 0.997 \\
		\cline{2-6}
		&  Failure & 0.997 & 0.981 & \textbf{0.996} & 0.988 \\
		\hline
		\multirow{3}{1.2cm}{\centering SISSA-L-A} &  Normal & \textbf{0.998} & \textbf{0.992} & \textbf{0.994} & \textbf{0.993} \\
		\cline{2-6}
		& Attack & \textbf{1.000} & \textbf{1.000} & \textbf{1.000} & \textbf{1.000} \\
		\cline{2-6}
		&  Failure & \textbf{0.998} & \textbf{0.992} & 0.992 & \textbf{0.992} \\
		\hline
		\hline
	\end{tabular}
	
    \begin{tablenotes}
	\footnotesize
	%\item[*] this is the ....  %此处加入注释*信息
	%\item[**] my website is ... %此处加入注释**信息
	\item Acc: Accuracy; Pre: Precision; F1: F1-measure 
\end{tablenotes}
\end{table}

\begin{table}
	\renewcommand{\arraystretch}{0.9}
	\centering
%	\captionsetup{font=Large} 
	\caption{The comparison of the  accuracy, precision, Recall and F1-score of each class under each SISSA model}
    \label{table: comparison of each class under each SISSA}
	\fontsize{8}{12}\selectfont
	\begin{tabular}{>{\centering\arraybackslash}m{1cm} | >{\centering\arraybackslash}m{1.2cm} | >{\centering\arraybackslash}m{0.9cm} | >{\centering\arraybackslash}m{0.9cm} | >{\centering\arraybackslash}m{0.9cm} | >{\centering\arraybackslash}m{0.9cm}}
		\hline
		\hline
		\multicolumn{1}{c|}{Model} & \multicolumn{1}{c|}{Class} & \multicolumn{1}{c|}{Acc} & \multicolumn{1}{c|}{Pre} & \multicolumn{1}{c|}{Recall} & \multicolumn{1}{c}{F1} \\
		\hline
		\multirow{7}{1cm}{\centering SISSA-C} & Normal & 0.832 & 0.403 & 0.361 & 0.381 \\
		\cline{2-6}
		& DDos & 0.971 & 0.913 & 0.880 & 0.896 \\
		\cline{2-6}
		& FI & 0.998 & \textbf{1.000} & 0.983 & 0.991 \\
		\cline{2-6}
		& FS & 0.863 & 0.520 & 0.536 & 0.528 \\
		\cline{2-6}
		& ReqNoRes & 0.977 & 0.932 & 0.905 & 0.918 \\
		\cline{2-6}
		& ResNoReq & 0.993 & 0.962 & 0.988 & 0.975 \\
		\cline{2-6}
		& Failure & 0.826 & 0.404 & 0.456 & 0.428 \\
		\hline
		\multirow{7}{1cm}{\centering SISSA-C-A} & Normal & 0.829 & 0.395 & 0.375 & 0.385 \\
		\cline{2-6}
		& DDos & 0.971 & 0.913 & 0.880 & 0.896 \\
		\cline{2-6}
		& FI & 0.998 & \textbf{1.000} & 0.988 & 0.994 \\
		\cline{2-6}
		& FS & 0.862 & 0.515 & 0.528 & 0.522 \\
		\cline{2-6}
		& ReqNoRes & 0.974 & 0.903 & 0.920 & 0.912 \\
		\cline{2-6}
		& ResNoReq & 0.991 & 0.971 & 0.967 & 0.969 \\
		\cline{2-6}
		& Failure & 0.833 & 0.421 & 0.447 & 0.434 \\
		\hline
		\multirow{7}{1cm}{\centering SISSA-R} & Normal & 0.961 & 0.853 & 0.880 & 0.866 \\
		\cline{2-6}
		& DDos & 0.992 & 0.966 & 0.981 & 0.973 \\
		\cline{2-6}
		& FI & \textbf{1.000} & \textbf{1.000} & \textbf{1.000} & \textbf{1.000} \\
		\cline{2-6}
		& FS & 0.978 & 0.934 & 0.913 & 0.923 \\
		\cline{2-6}
		& ReqNoRes & 0.994 & 0.982 & 0.979 & 0.981 \\
		\cline{2-6}
		& ResNoReq & 0.999 & 0.996 & 0.998 & 0.997 \\
		\cline{2-6}
		& Failure & 0.978 & 0.933 & 0.913 & 0.922 \\
		\hline
		\multirow{7}{1cm}{\centering SISSA-R-A} & Normal & 0.957 & 0.849 & 0.849 & 0.849 \\
		\cline{2-6}
		& DDos & 0.991 & 0.962 & 0.979 & 0.970 \\
		\cline{2-6}
		& FI & \textbf{1.000} & \textbf{1.000} & \textbf{1.000} & \textbf{1.000} \\
		\cline{2-6}
		& FS & 0.978 & 0.931 & 0.915 & 0.923 \\
		\cline{2-6}
		& ReqNoRes & 0.994 & 0.984 & 0.975 & 0.980 \\
		\cline{2-6}
		& ResNoReq & 0.999 & 0.996 & \textbf{1.000} & 0.998 \\
		\cline{2-6}
		& Failure & 0.975 & 0.909 & 0.915 & 0.912 \\
		\hline
		\multirow{7}{1cm}{\centering SISSA-L} & Normal & 0.994 & 0.984 & 0.971 & 0.978 \\
		\cline{2-6}
		& DDos & 0.998 & 0.996 & 0.992 & 0.994 \\
		\cline{2-6}
		& FI & \textbf{1.000} & \textbf{1.000} & \textbf{1.000} & \textbf{1.000} \\
		\cline{2-6}
		& FS & 0.996 & 0.988 & 0.984 & 0.986 \\
		\cline{2-6}
		& ReqNoRes & 0.999 & 0.994 & \textbf{0.998} & 0.996 \\
		\cline{2-6}
		& ResNoReq & 1.000 & 0.998 & \textbf{1.000} & 0.999 \\
		\cline{2-6}
		& Failure & 0.997 & 0.981 & \textbf{0.996} & 0.988 \\
		\hline
		\multirow{7}{1cm}{\centering SISSA-L-A} & Normal & \textbf{0.998} & \textbf{0.992} & \textbf{0.994} & \textbf{0.993} \\
		\cline{2-6}
		& DDos & \textbf{0.999} & \textbf{0.996} & \textbf{0.998} & \textbf{0.997} \\
		\cline{2-6}
		& FI & \textbf{1.000} & \textbf{1.000} & \textbf{1.000} & \textbf{1.000} \\
		\cline{2-6}
		& FS & \textbf{0.999} & \textbf{1.000} & \textbf{0.996} & \textbf{0.998} \\
		\cline{2-6}
		& ReqNoRes & \textbf{0.999} & \textbf{0.998} & \textbf{0.998} & \textbf{0.998} \\
		\cline{2-6}
		& ResNoReq & \textbf{1.000} & \textbf{1.000} & \textbf{1.000} & \textbf{1.000} \\
		\cline{2-6}
		& Failure & \textbf{0.998} & \textbf{0.992} & 0.992 & 0.992 \\
		\hline
		\hline
	\end{tabular}

    \begin{tablenotes}
	\footnotesize
	%\item[*] this is the ....  %此处加入注释*信息
	%\item[**] my website is ... %此处加入注释**信息
	\item Acc: Accuracy; Pre: Precision; F1: F1-measure; 
	\item DDos: Distributed Denial of Service; FI: Fake Interface;
	\item FS: Fake Source; Distributed Denial of Service;
	\item ReqNoRes: Request without Response; 
	\item ResNoReq: Response without Request
\end{tablenotes}

\end{table}

\begin{table*}
	\renewcommand{\arraystretch}{0.9}
	\centering
	%	\captionsetup{font=Large} 
	\caption{SISSA's characteristics \& computational resources }
	\label{table: SISSA's characteristics} 
	\fontsize{8}{12}\selectfont
	\begin{tabular}{>{\centering\arraybackslash}m{1.3cm} | >{\centering\arraybackslash}m{2cm} | >{\centering\arraybackslash}m{1.5cm} | >{\centering\arraybackslash}m{1.5cm} | >{\centering\arraybackslash}m{1.5cm} | >{\centering\arraybackslash}m{2.2cm} | >{\centering\arraybackslash}m{1cm} | >{\centering\arraybackslash}m{1.8cm}}
		\hline
		\hline
		\multirow{2}{*}{Windows}  & \multirow{2}{*}{Model} &  \multirow{2}{*}{\thead{Total \\Params}} &  \multirow{2}{*}{\thead{Params \\size (MB)}} & \multirow{2}{*}{\thead{Input \\size (MB)}} & \multirow{2}{*}{\thead{Forward/Backward \\Pass size (MB)}} & \multirow{2}{*}{\thead{Total \\size}} & \multirow{2}{*}{\thead{Time Cost (s)}} \\
		& & & & & & & \\
		\hline
		\multirow{6}{*}{w32} & SISSA-C & 73352 & 0.29 & 0.01 & 1 & 1.29 & 0.0005 \\
		\cline{2-8}
		& SISSA-C-A & 74168 & 0.3 & 0.01 & 1 & 1.3 & 0.0007 \\
		\cline{2-8}
		& SISSA-R & 579911 & 2.32 & 0.01 & 0.07 & 2.4 & 0.0005 \\
		\cline{2-8}
		& SISSA-R-A & 629447 & 2.52 & 0.01 & 0.17 & 2.69 & 0.0007 \\
		\cline{2-8}
		& SISSA-L & 741191 & 2.96 & 0.01 & 0.07 & 3.04 & 0.0005 \\
		\cline{2-8}
		& SISSA-L-A & 790727 & 3.16 & 0.01 & 0.17 & 3.34 & 0.0007 \\
		\hline
		\multirow{6}{*}{w48} & SISSA-C & 73768 & 0.3 & 0.01 & 1.01 & 1.31 & 0.0005 \\
		\cline{2-8}
		& SISSA-C-A & 74584 & 0.3 & 0.01 & 1.01 & 1.31 & 0.0007 \\
		\cline{2-8}
		& SISSA-R & 844519 & 3.38 & 0.01 & 0.12 & 3.5 & 0.0005 \\
		\cline{2-8}
		& SISSA-R-A & 894055 & 3.58 & 0.01 & 0.27 & 3.85 & 0.0007 \\
		\cline{2-8}
		& SISSA-L & 1011943 & 4.05 & 0.01 & 0.12 & 4.17 & 0.0005 \\
		\cline{2-8}
		& SISSA-L-A & 1061479 & 4.25 & 0.01 & 0.27 & 4.52 & 0.0007 \\
		\hline
		\multirow{6}{*}{w64} & SISSA-C & 74184 & 0.3 & 0.01 & 1.02 & 1.32 & 0.0005 \\
		\cline{2-8}
		& SISSA-C-A & 75000 & 0.3 & 0.01 & 1.91 & 2.21 & 0.0007 \\
		\cline{2-8}
		& SISSA-R & 1109127 & 4.44 & 0.01 & 0.16 & 4.61 & 0.0005 \\
		\cline{2-8}
		& SISSA-R-A & 1158663 & 4.63 & 0.01 & 0.36 & 5 & 0.0008 \\
		\cline{2-8}
		& SISSA-L & 1282695 & 5.13 & 0.01 & 0.36 & 5.7 & 0.0008 \\
		\cline{2-8}
		& SISSA-L-A & 1332231 & 5.33 & 0.01 & 0.36 & 5.7 & 0.0008 \\
		\hline
		\multirow{6}{*}{w80} & SISSA-C & 74600 & 0.3 & 0.01 & 1.04 & 1.34 & 0.0005 \\
		\cline{2-8}
		& SISSA-C-A & 75416 & 0.3 & 0.01 & 1.04 & 1.35 & 0.0007 \\
		\cline{2-8}
		& SISSA-R & 1373735 & 5.49 & 0.01 & 0.22 & 5.72 & 0.0005 \\
		\cline{2-8}
		& SISSA-R-A & 1423271 & 5.69 & 0.01 & 0.46 & 6.16 & 0.0008 \\
		\cline{2-8}
		& SISSA-L & 1553447 & 6.21 & 0.01 & 0.22 & 6.44 & 0.0006 \\
		\cline{2-8}
		& SISSA-L-A & 1602983 & 6.41 & 0.01 & 0.46 & 6.88 & 0.0008 \\
		\hline
		\multirow{6}{*}{w96} & SISSA-C & 75016 & 0.3 & 0.01 & 1.06 & 1.37 & 0.0005 \\
		\cline{2-8}
		& SISSA-C-A & 75832 & 0.3 & 0.01 & 1.07 & 1.38 & 0.0008 \\
		\cline{2-8}
		& SISSA-R & 1638343 & 6.55 & 0.01 & 0.27 & 6.83 & 0.0005 \\
		\cline{2-8}
		& SISSA-R-A & 1687879 & 6.75 & 0.01 & 0.57 & 7.33 & 0.0008 \\
		\cline{2-8}
		& SISSA-L & 1824199 & 7.3 & 0.01 & 0.27 & 7.58 & 0.0006 \\
		\cline{2-8}
		& SISSA-L-A & 1873735 & 7.49 & 0.01 & 0.57 & 8.07 & 0.0008 \\
		\hline
		\multirow{6}{*}{w112} & SISSA-C & 75432 & 0.3 & 0.01 & 1.09 & 1.4 & 0.0005 \\
		\cline{2-8}
		& SISSA-C-A & 76248 & 0.3 & 0.01 & 1.09 & 1.41 & 0.0008 \\
		\cline{2-8}
		& SISSA-R & 1902951 & 7.61 & 0.01 & 0.33 & 7.95 & 0.0005 \\
		\cline{2-8}
		& SISSA-R-A & 1952487 & 7.81 & 0.01 & 0.67 & 8.5 & 0.0008 \\
		\cline{2-8}
		& SISSA-L & 2094951 & 8.38 & 0.01 & 0.33 & 8.72 & 0.0006 \\
		\cline{2-8}
		& SISSA-L-A & 2144487 & 8.58 & 0.01 & 0.67 & 9.26 & 0.0008 \\
		\hline
		\multirow{6}{*}{w128} & SISSA-C & 75848 & 0.3 & 0.01 & 1.12 & 1.43 & 0.0005 \\
		\cline{2-8}
		& SISSA-C-A & 76664 & 0.31 & 0.01 & 1.12 & 1.44 & 0.0007 \\
		\cline{2-8}
		& SISSA-R & 2167559 & 8.67 & 0.01 & 0.39 & 9.08 & 0.0006 \\
		\cline{2-8}
		& SISSA-R-A & 2217095 & 8.87 & 0.01 & 0.79 & 9.67 & 0.0008 \\
		\cline{2-8}
		& SISSA-L & 2365703 & 9.46 & 0.01 & 0.39 & 9.87 & 0.0006 \\
		\cline{2-8}
		& SISSA-L-A & 2415239 & 9.66 & 0.01 & 0.79 & 10.46 & 0.0008 \\
		\hline
		\hline
	\end{tabular}
\end{table*}

Figure \ref{fig: ROC-AUC} shows the ROC curves and AUC values for six models with different backbones. It can be observed that the AUC values of LSTM based models (SISSA-L and SISSA-L-A) are range from a minimum of 0.93, with the majority approaching 1. This indicates that LSTM-based models can effectively distinguish between each type of attack, demonstrating superior performance.

 In stark contrast, the SISSA-C and SISSA-C-A models illustrated in Figure \ref{fig: ROC-AUC}(a) and (b) exhibit a distinct pattern. The ROC curves associated with these models deviate from the typical convergence towards the upper-left corner. Simultaneously, the AUC values they entail indicate challenges in accurately discerning various attacks, resulting in reduced predictive accuracy.

%In Figure \ref{fig: ROC-AUC}, we present the ROC curves and AUC values for six models with different backbones across seven classification tasks. It can be observed that in the results of LSTM based models (SISSA-L and SISSA-L-A), The AUC values for each class are in proximity to or equal to 1. This indicates that LSTM can effectively distinguish between each type of attack, demonstrating superior performance.
%
%The ROC curves and corresponding AUC values for six models in a multi-classification scenario are illustrated in Figures \ref{fig: ROC-AUC} (a)-(f). We have zoomed in on the upper-left region of Figures \ref{fig: ROC-AUC} (f) to emphasize distinctions among the ROC curves for different classifications. For the LSTM based model, Figures \ref{fig: ROC-AUC} (e) showcases its ROC curve and AUC values. The AUC values for various classifications range from a minimum of 0.93, with the majority approaching 1, indicating a commendable predictive accuracy. Notably, Figures \ref{fig: ROC-AUC} (f) demonstrates that the LSTM model augmented with RSA exhibits superior performance, with AUC values predominantly reaching 1.

 In the case of Figure \ref{fig: ROC-AUC} (c) and (d), the ROC curves for the RNN-based models tend to converge towards the upper-left corner, suggesting commendable predictive accuracy. However, its performance falls slightly short of the LSTM-based models. These observations collectively underscore the efficacy of the models and highlight the impact of incorporating RSA, particularly evident in the enhanced predictive capabilities of the LSTM-based model.
 
 Figure \ref{fig: Wsize} delineates the comparative performance of six distinct models as a function of varying window sizes. In the conducted experiments, the SISSA-L-A and SISSA-L models consistently outperformed their counterparts across all four measured metrics. It was observed that with an increase in window height, the RNN-backed models SISSA-R and SISSA-R-A exhibited a marked improvement in all indices. Conversely, models predicated on a CNN-based backbone did not demonstrate significant alterations in performance. This phenomenon can be attributed to the RNN and LSTM's inherent capability to exploit temporal information provided by an enlarged window, thereby enhancing the model's performance. In contrast, the SISSA-C and SISSA-C-L models appear to be less adept at capturing temporal characteristics and the inter-message associations, leading to a stagnant performance outcome. Additionally, it is noted that the integration of a RSA generally enhances the model's overall metrics. However, it necessitates a more extended training duration to reach optimal efficacy, compared to its counterparts devoid of the RSA.

 Table \ref{table: comparison of each SISSA } presents a comprehensive evaluation of six distinct SISSA models across three categorical scenarios: normal message windows, malicious attack message windows, and random hardware failure message windows. The table is formatted to emphasize the highest metric values for each category by employing bold typeface. This accentuation reveals that the SISSA-L-A model exhibits superior performance, achieving near-perfect accuracy in the identification of malicious attack message windows. It is noteworthy, however, that during the detection of random hardware failure message windows, the recall metric for SISSA-L-A is marginally eclipsed by that of SISSA-L. This suggests a propensity of the SISSA-L-A model to occasionally misclassify normal operational modes as faulty. This phenomenon may be attributed to a degree of overfitting engendered by the RSA within the model's architecture, underscoring a potential area for further refinement. 
 
 Table \ref{table: comparison of each class under each SISSA} elucidates the performance of each SISSA model across seven classification categories, focusing on the metrics of accuracy, precision, recall, and F1-score. The results highlight that, given a sufficient training set, the detection rate for Fake Interface attacks by all six proposed models is near perfect. The models exhibit varying degrees of efficacy in identifying other forms of attacks and random hardware failures, which is reflective of their differential capabilities in temporal feature extraction. Notably, the SISSA-L-A model stands out with the most robust performance, demonstrating high accuracy in the recognition of diverse attack vectors and random functional failures. This suggests that the synergy between the LSTM backbone and RSA is particularly adept at discerning between multiple attack patterns and hardware malfunctions within multi-node SOME/IP communication.
   
\subsubsection{Overhead}

Table \ref{table: SISSA's characteristics} presents a comprehensive analysis of parameters and time costs associated with six distinct models, comparing them across various window sizes. It is observed that an increase in window height results in a slower detection speed, attributed to the simultaneous increase in the number of parameters. When comparing models with identical window values, those incorporating RSA tend to exhibit higher parameter counts and slower detection times compared to counterparts without RSA. Furthermore, notable differences in parameter quantities exist among different model architectures. In particular, the SISSA-L model exhibits a higher parameter count and longer detection time in comparison to SISSA-R, the latter surpassing SISSA-C in both aspects. Meanwhile, the graphics memory increases with the growth of window height. Considering that the resolution of message data itself is not substantial, memory usage is unlikely to pose a hindrance  in practical applications. It's noteworthy to mention that detection time of SISSA is less than 1ms, which can meet  in-vehicle network traffic and real-time requirements defined in IEEE 802.1DG \cite{IEEE8021DG}.

\section{Discussion}
%\subsection{Real-world application}
In the experimental evaluation, we observed distinct variations in detection rates and speeds among the six SISSA models that developed for identifying cyber attacks and random hardware failures. Owing to the comprehensive Python code framework, SISSA can be seamlessly integrated as a middleware within automotive communication networks to facilitate real-time monitoring of SOMP/IP network activities. If this middleware operates in a strictly localized environment, the SISSA network itself remains safeguarded against external attacks. Conversely, when connected to cloud services, the model's efficiency can be incrementally enhanced in alignment with vehicular modifications, allowing for the transmission of anomalous states back to remote data centers. Additionally, the versatility of SISSA is evident in its capacity to adapt to varying packet traffic types through modifications in its structural design and the dateset used for training.

In practical applications, critical factors such as the model's detection rate, inference time, and memory usage must be meticulously considered. Tailoring the window height and model structure to suit the complexity of communication scenarios is imperative. For instance, in our specific use-case, employing the SISSA-L-A model with an LSTM backbone and a monitoring window of 128 consecutive packets, we achieved a remarkable 99.7\% accuracy across seven classification challenges. Concurrently, the model's inference time aligns with the prerequisites for real-time monitoring. However, it should be noted that increasing the window height further may yield diminishing returns in accuracy relative to the augmented inference time.

Furthermore, it is worth noting that the low-dimensional nature of the message data results in a minimal GPU memory footprint during inference. This aspect underscores the feasibility of running SISSA for safety and security analyses in automotive systems without significantly impacting other GPU-based tasks. Such an attribute is vital for the practical deployment of the application.
%\subsection{Limitations}

\section{Conclusion}
\label{Conclusion}
In this work, we propose a novel approach, termed SISSA, to monitoring the safety and security of 
ECUs in connected automated vehicles using SOME/IP communication traffic. SISSA employs Weibull distribution to model random hardware failures and identifies five classic cyberattacks. Leveraging three deep learning models with residual self-attention mechanisms, SISSA effectively extracts features from SOME/IP packets to distinguish between normal ECU operation, functional failures, and cyberattacks. We have curated a comprehensive dataset encompassing various classes for evaluation, a noteworthy contribution given the scarcity of publicly available datasets with similar characteristics. The experimental results indicate SISSA's high effectiveness, achieving an average F1-score of 99.8\% for malfunction identification and a perfect average F1-score of 100\% for cyberattack detection. 
The detection speed and model overhead meet the communication needs of automotive networks while also accommodating the computational resource requirements of ECUs.

In the future, we intend to expand the scope of our framework to encompass additional in-vehicle communication protocols, including the well-established CAN bus. Besides, modeling cyberattacks on in-vehicle networks emerges as an intriguing and worthwhile avenue for exploration.

\bibliographystyle{IEEEtran}
\bibliography{IEEEabrv,reference}

\begin{comment}

\end{comment}

\newpage

\begin{comment}

\section{Biography Section}
If you have an EPS/PDF photo (graphicx package needed), extra braces are
 needed around the contents of the optional argument to biography to prevent
 the LaTeX parser from getting confused when it sees the complicated
 $\backslash${\tt{includegraphics}} command within an optional argument. (You can create
 your own custom macro containing the $\backslash${\tt{includegraphics}} command to make things
 simpler here.)
 
\vspace{11pt}

\bf{If you include a photo:}\vspace{-33pt}
\begin{IEEEbiography}[{\includegraphics[width=1in,height=1.25in,clip,keepaspectratio]{fig1}}]{Michael Shell}
Use $\backslash${\tt{begin\{IEEEbiography\}}} and then for the 1st argument use $\backslash${\tt{includegraphics}} to declare and link the author photo.
Use the author name as the 3rd argument followed by the biography text.
\end{IEEEbiography}

\vspace{11pt}

\bf{If you will not include a photo:}\vspace{-33pt}
\begin{IEEEbiographynophoto}{John Doe}
Use $\backslash${\tt{begin\{IEEEbiographynophoto\}}} and the author name as the argument followed by the biography text.
\end{IEEEbiographynophoto}

\vfill
\end{comment}

\end{document}